\documentclass{jfm}

\usepackage{afterpage}
\usepackage{longtable}
\usepackage[english]{babel}
\usepackage{dcolumn}
\usepackage{bm}
\usepackage{pifont}
\usepackage{amsfonts}
\usepackage{amsmath}
\usepackage{upmath}
\usepackage{amssymb}
\usepackage{wasysym}
\usepackage{amsbsy}
\usepackage{graphicx}
\usepackage{natbib}
\usepackage{psfrag}
\usepackage{lineno}
\usepackage{color}
\usepackage{multirow}
\usepackage{cases}
\usepackage{stackengine}
\usepackage{float}
\usepackage{blkarray}
\usepackage{cancel}
\usepackage{mathrsfs}
\usepackage{empheq}

\providecommand\bnabla{\boldsymbol{\nabla}}
\providecommand\bcdot{\boldsymbol{\cdot}}
\newcommand{\boldm}[1]{\boldsymbol{#1}}

\newcommand{\Oh}{\mbox{\textit{Oh}}} 		
\newcommand{\Reycap}{\mbox{$\mathscr{R}$}} 	
\newcommand{\Bou}{\mbox{$\mathscr{B}$}} 	

\title[Temporal stability of free liquid threads with surface viscoelasticity]%
{Temporal stability of free liquid threads with surface viscoelasticity}

\author%
[A. Mart\'inez-Calvo, A. Sevilla]%
{A. Mart\'inez-Calvo\thanks{Email address for correspondence: amcalvo@ing.uc3m.es}\ns and A. Sevilla}

\affiliation{%
Grupo de Mec\'anica de Fluidos,
Departamento de Ingenier\'ia T\'ermica y de Fluidos,
Universidad Carlos III de Madrid,
Av.~Universidad 30,
28911 Legan\'es (Madrid),
Spain}

\begin{document}
\maketitle

\begin{abstract}

We analyse the effect of surface viscoelasticity on the temporal stability of a free cylindrical liquid jet coated with insoluble surfactant, extending the results of~\citet{Timmermans02}. Our development requires, in particular, deriving the correct expressions for the normal and tangential stress boundary conditions at a general axisymmetric interface when surface viscosity is modelled with the Boussinesq--Scriven constitutive equation. These stress conditions are applied to obtain a new dispersion relation for the liquid thread, which is solved to describe its temporal stability as a function of four governing parameters, namely the capillary Reynolds number, the elasticity parameter, and the shear and dilatational Boussinesq numbers. It is shown that both surface viscosities have a stabilising influence for all values of the capillary Reynolds number and elasticity parameter, the effect being more pronounced at low capillary Reynolds numbers. The wavenumber of maximum amplification depends non-monotonically on the Boussinesq numbers, especially for very viscous threads at low values of the elasticity parameter. Finally, two different lubrication approximations of the equations of motion are derived. While the validity of the leading-order model is limited to small enough values of the elasticity parameter and of the Boussinesq numbers, a higher-order parabolic model is able to accurately capture the linearised behaviour for the whole range of values of the four control parameters.

\end{abstract}

\begin{keywords}
Capillary flows, Instability
\end{keywords}

\section{Introduction \label{sec:intro}}

It has been known for a long time that the presence of surface-active molecules modifies the dynamics of fluid interfaces -- for a short historical account the reader is referred to~\citet{ScrivenNature}. The crucial importance of surfactants in the respiratory system~\citep{VanGolde88}, as well as in many applications like the generation and stabilisation of emulsions, foams and contrast agents for medical imaging~\citep{Rodriguez2015}, justifies the intense research effort oriented towards a quantitative understanding of their effects on the dynamics of interfaces. One important effect associated with adsorbed surfactant molecules is their resistance to surface deformation, which manifests macroscopically as an effective interfacial rheology~\citep{LangevinARFM}. Experiments and theory have clearly demonstrated that surface viscous effects play a central role in many interfacial flows like foams~\citep{Joye94}, liquid films~\citep{Scheid10}, liquid bridges~\citep{Ponce16b,Ponce16a} and drop breakup~\citep{Ponce17}, to cite a few. In particular, the conclusion of~\citet{Ponce17} that surface viscosity dictates the amount of surfactant present in the satellite droplets formed after pinch-off constitutes the main motivation of the present work.

Similarly, the effects of surface elasticity induced by the presence of surfactants have been studied in several configurations, e.g. liquid threads~\citep{Craster2002,Timmermans02}, dip coating~\citep{Campana2011, Seiwert2014, Champougny2015}, bubbles immersed in a viscous medium~\citep{Hameed2008} or drop deformation and breakup~\citep{StoneLeal1990, Milliken1993}. 

Previous related studies have extended the seminal work of~\citet{Tomotika} on the stability of liquid threads embedded in an immiscible unbounded liquid by including the effects of surfactants. In particular,~\citet{Hajiloo87} considered surface viscosities in the formulation but they did not account for surface tension gradients. The work of~\citet{Hansen99} deals with soluble surfactants and also accounts for surface diffusion and Marangoni stresses, but does not include the effect of surface viscosities.~\citet{Kwak01} extended the problem for an insoluble surfactant by including internal and external coaxial solid boundaries, and also retaining the effect of surface diffusion, but not the effect of surface viscosities. Previously,~\citet{Carroll1974} analysed both experimentally and theoretically the effect of solid oleophilic filaments coated by a cylindrical oil film, observing that the presence of an insoluble surfactant at the outer oil-air interface decreases the growth rate of instabilities substantially. These authors also studied the nonlinear dynamics of drop growth and breakup, and contemplated the solid-oil-water configuration, but they did not account for variations in the surface tension coefficient, nor surface shear viscosity, including only the effect of dilatational viscosity.

The nonlinear dynamics of liquid threads in the presence of insoluble and soluble surfactants has also been studied by means of experiments~\citep{Roche2009}, theory~\citep{Timmermans02} and simulations~\citep{Dravid2006,Campana2006}. In particular, there are several works devoted to analyse the nonlinear dynamics by developing one-dimensional (1D) long-wave approximations of the Navier-Stokes equations, thereby reducing the computational cost and facilitating analytical developments. Following the same procedure as~\citet{EggersDupont} and~\citet{GyC} for a clean interface, several authors have derived leading-order 1D models accounting for the presence of surfactant; for instance~\citet{Ambra1999} study the effect of insoluble surfactant on the breakup of liquid bridges, and~\citet{Craster2002} and~\citet{Craster2009} examine the capillary breakup of liquid jets. However, as pointed out by~\citet{Timmermans02}, higher-order models are generally needed since the leading-order equations fail at describing the linear stage of the Plateau-Rayleigh instability when the elasticity parameter is sufficiently large. It is important to point out that, to date, no attempt has been made to develop a nonlinear 1D model that accounts for surface viscous effects, which constitutes one of the objectives of the present work.

To the best of our knowledge, the only study of the effect of surface viscosity on the Plateau-Rayleigh instability of a free liquid jet~\citep{Plateau,Rayleigh1} is due to~\citet{Whitaker76} who, as pointed out by~\citet{Hansen99} and by~\citet{Timmermans02}, deduced an incorrect dispersion relation. The work of~\citet{Whitaker76} also includes the effect of solubility in the dispersion relation, although it is neither analysed nor discussed. Besides, the results and conclusions of~\citet{Whitaker76} are restricted to the limit of dominant inertia.

Therefore, in the present work we extend the results of~\citet{Timmermans02} by including the effect of shear and dilatational surface viscosities in the Boussinesq--Scriven approximation~\citep{Boussinesq1913,Scriven60}, examining their influence in the linear dispersion relation and also in the maximum temporal growth rate and its corresponding wavenumber. Besides, since the literature is spotted with diverse mistakes and missing terms in the normal and tangential stress balances at a general axisymmetric interface when surface viscosities are considered, here we deduce both boundary conditions in detail. We also derive a leading-order and a parabolic set of nonlinear 1D equations accounting for Marangoni stresses and surface viscosities, showing that only the higher-order parabolic approximation is able to accurately reproduce the exact linear dispersion relation when the values of the surface shear and dilatational viscosities are large enough.

The remainder of the paper is organised as follows. In~\S\ref{sec:formulation} we derive the interfacial stress boundary conditions needed to study axisymmetric free surface flows in the presence of insoluble viscous monolayers, correcting errors of previous works. The formalism is then applied in \S\ref{sec:results} to study the canonical case of the capillary instability of a liquid cylinder, focusing on the effect of surface viscosity on the temporal amplification of small disturbances. In~\S\ref{sec:longwave} we derive the 1D leading-order and parabolic sets of equations including Marangoni stresses and surface viscosities, and compare their associated dispersion relations with that derived in~\S\ref{sec:results}. Conclusions are drawn in~\S\ref{sec:conclusions}.

\section{Stress boundary conditions at an axisymmetric viscoelastic interface \label{sec:formulation}}

\begin{figure}
    \centering
    \includegraphics[width=0.9\textwidth]{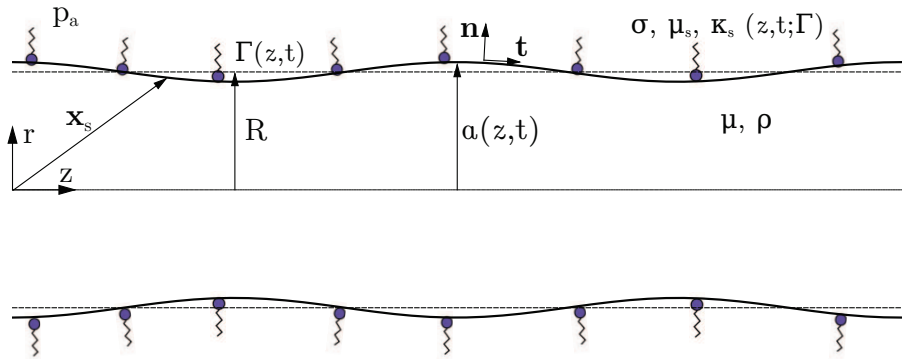}
    \caption{Sketch of the flow configuration.\label{fig:figure1}}
\end{figure}

Let us consider the axisymmetric interface between an incompressible liquid of density $\rho$ and viscosity $\mu$ and a passive surrounding gaseous atmosphere at uniform pressure $p_a$. The interface is coated with a surface concentration of insoluble surfactant molecules, $\Gamma(z,t)$, and a cylindrical coordinate system is adopted to describe the flow, as depicted in figure~\ref{fig:figure1}. Note that $r$, $z$ and $t$ stand for the radial and axial coordinates and time, respectively. Hereinafter, every surface quantity will be denoted by the subscript $s$, whereas the tangential and normal projections will be referred to with the subscripts $t$ and $n$, respectively.

The role of the surfactant is twofold. First, it reduces the surface tension coefficient, $\sigma$, by an amount that depends on $\Gamma$. Second, it induces an effective surface rheology through elastic and viscous stresses. The former are the Marangoni stresses generated by the imbalances of $\sigma$ produced by the variations of $\Gamma$ along the interface, which leads to the surface elasticity. The surface viscous stresses are usually described through two surface viscosity coefficients, namely the surface shear and dilatational viscosities, $\mu_s(\Gamma)$ and $\kappa_s(\Gamma)$, respectively. Hence, the liquid-gas interface of radius $r = a(z,t)$ is described as a compressible two-dimensional Newtonian surface with negligible surface density and obeying the Boussinesq--Scriven constitutive equation. To derive the latter equation the free surface is parametrised in terms of the axial coordinate $z$ and the azimuthal angle $\theta$ as
\begin{equation}
\boldm{x}_s = \boldm{x}_s(z,\theta,t) = a(z,t) \, \boldm{e}_r + z \, \boldm{e}_z,
\end{equation}
where $\boldm{x}_s$ is the position vector of an arbitrary point lying on the surface, and $\boldm{e}_r$ and $\boldm{e}_z$ are the unit radial and axial vectors, respectively. For the following development surface operators need to be defined. To that end we build an orthonormal curvilinear basis $\{\boldm{t}, \boldm{e}_{\theta}, \boldm{n} \}$ intrinsic to the surface, where
\begin{equation}
\boldm{n}  = \frac{1}{\sqrt{1 + a'^2}} (\boldm{e}_r - a' \, \boldm{e}_z), \, \, \, \boldm{t} =  \frac{1}{\sqrt{1 + a'^2}} (a' \, \boldm{e}_r + \boldm{e}_z),
\end{equation}
with $\boldm{n}$ the unit normal vector and $\boldm{t}$, $\boldm{e}_{\theta}$ the unit meridional and azimuthal tangent vectors. Throughout the paper, primed variables denote their partial derivatives with respect to $z$. The latter basis introduces a covariant metric tensor of components $\mathsfi{g}_{11} = 1 + (a')^2$, $\mathsfi{g}_{22} = a^2$, $\mathsfi{g}_{12} = \mathsfi{g}_{21} = 0$, whose determinant is given by $g = a^2[1+(a')^2]$. Hence, the surface gradient operator, $\bnabla_s$, is defined as
\begin{equation}
\bnabla_s = \frac{1}{\sqrt{1 + a'^2}} \boldm{t} \frac{\partial}{\partial z} + \frac{1}{a} \boldm{e}_{\theta} \frac{\partial}{\partial \theta}. \label{eq:surfacegradient}
\end{equation}
The velocity field of the fluid at the surface, $\boldm{u}_s$, can be decomposed into a tangential velocity $\boldm{u}_t$, and a normal velocity $\boldm{u}_n$,
\begin{equation}
\boldm{u}_s  = \boldm{u}(\boldm{x}_s,t)= \boldm{u}_t(\boldm{x}_s,t) + \boldm{u}_n(\boldm{x}_s,t) = \boldm{u}_t(\boldm{x}_s,t) + u_n(\boldm{x}_s,t) \,\boldm{n},
\end{equation}
where $\boldm{u}_t = \mathsfbi{I}_s \bcdot \boldm{u}_s$, $\boldm{u}_n  = \boldm{n}\boldm{n} \bcdot \boldm{u}_s$, $\mathsfbi{I}_s = \mathsfbi{I} - \boldm{n}\boldm{n}$ is the surface projection operator and $\mathsfbi{I}$ is the three-dimensional identity tensor. Note that the surface gradient operator given by equation~\eqref{eq:surfacegradient} can also be written as $\bnabla_s = \mathsfbi{I}_s\bcdot \bnabla$ in terms of the standard gradient and the surface projection operator. Since the interface is axisymmetric, $\boldm{u}_t(\boldm{x}_s,t) = u_t(\boldm{x}_s,t) \, \boldm{t}$, and thus the normal and the meridional tangent velocity components read, respectively,
\begin{equation}
u_n = \frac{u - a' w}{\sqrt{1+a'^2}}, \, \, \,  u_t = \frac{w + a' u}{\sqrt{1+a'^2}},\label{eq:un_ut}
\end{equation}
where $u$ and $w$ are the radial and axial components of the fluid velocity respectively. For simplicity, in equation~\eqref{eq:un_ut} we have omitted the evaluation of $u$ and $w$ at $r = a(z,t)$.

To derive the stress balance for the fluid interface we apply the integral momentum conservation equation to a surface element $S$, $\int \int_S ( \hat{\mathsfbi{T}}-\mathsfbi{T} ) \bcdot \boldm{n} \, \mathrm{d}A + \int_C \mathsfbi{T}_s \bcdot \boldm{n}_l \, \mathrm{d}l = 0$, where $\hat{\mathsfbi{T}}$ and $\mathsfbi{T}$ are the stress tensors of the outer and inner fluids evaluated at the interface, respectively, and $\mathsfbi{T}_s$ is their superficial counterpart, which is proportional to the line element enclosing the surface, $C$, whose unitary normal vector embedded in $S$ is $\boldm{n}_l$, and thus $\boldm{n} \bcdot \boldm{n}_l = 0$. The generalised Stokes theorem is applied to the line integral to give $( \hat{\mathsfbi{T}}-\mathsfbi{T} ) \bcdot \boldm{n} + \bnabla_s \bcdot \mathsfbi{T}_s - (\bnabla_s \bcdot \boldm{n})\boldm{n} \bcdot \mathsfbi{T}_s = 0$~\citep[see appendix A of][]{RiveroScheid2018}. In the present work, the surface stress tensor $\mathsfbi{T}_s$ is modelled with the well-known Boussinesq--Scriven approximation~\citep{Boussinesq1913,Scriven60} in the limit of negligible surface density, which disregards complex interfacial rheology and assumes that the deviatoric component of $\mathsfbi{T}_s$ is isotropic, linear and instantaneous in the surface rate-of-strain tensor,
\begin{equation}\label{eq:surface_tensor}
\mathsfbi{T}_s =  \mathsfbi{I}_s \left[ \sigma + (\kappa_{s} -\mu_{s})(\bnabla_s \bcdot \boldm{u}_s) \right] + \mu_s \left[ (\bnabla_s \boldm{u}_s) \bcdot \mathsfbi{I}_s + \mathsfbi{I}_s \bcdot (\bnabla_s \boldm{u}_s)^T\right].
\end{equation}
According to~\eqref{eq:surface_tensor}, $\mathsfbi{T}_s$ is restricted to be tangent to the interface, i.e. $\boldm{n} \bcdot \mathsfbi{T}_s = 0$, and thus the interfacial stress balance in differential form simplifies to
\begin{equation}
( \hat{\mathsfbi{T}}-\mathsfbi{T} ) \bcdot \boldm{n} + \bnabla_s \bcdot \mathsfbi{T}_s = 0.
\label{eq:surfdiff}
\end{equation}
Finally, introducing~\eqref{eq:surface_tensor} for $\mathsfbi{T}_s$ into~\eqref{eq:surfdiff}, the surface equation of motion reads
\begin{align}\label{eq:eqsurface}
&(\hat{\mathsfbi{T}}-\mathsfbi{T}) \bcdot \boldm{n} + \bnabla_s \sigma - \boldm{n}(\bnabla_s \bcdot \boldm{n}) \sigma + \bnabla_s \left[(\kappa_s-\mu_s) (\bnabla_s \bcdot \boldm{u}_s)\right]  \nonumber & \\
& -\boldm{n} (\bnabla_s \bcdot \boldm{n}) (\kappa_s - \mu_s) (\bnabla_s \bcdot \boldm{u}_s) + \bnabla_s \bcdot \left\{ \mu_s \left[(\bnabla_s \boldm{u}_s) \bcdot \mathsfbi{I}_s +\mathsfbi{I}_s \bcdot (\bnabla_s \boldm{u}_s)^{\text{T}} \right]\right\} = 0,
\end{align}
which is Newton's second law for a fluid surface with negligible surface density. Since the flow is considered axisymmetric,~\eqref{eq:eqsurface} results in two boundary conditions at the free surface. Hence, taking the inner product of~\eqref{eq:eqsurface} with $\boldm{n}$ and $\boldm{t}$, and considering that the outer fluid remains at constant pressure, $\hat{\mathsfbi{T}} = -p_a \mathsfbi{I}$, the normal and tangential stress balances in cylindrical coordinates read, respectively,
\begin{align}
& \left(p \bigr \rvert_{r = a(z,t)} - p_a \right) - \frac{2 \mu}{1+a'^2} \left[ \frac{\partial u}{\partial r} + a'^2 \frac{\partial w}{\partial z} - a' \left(\frac{\partial w}{\partial r} + \frac{\partial u}{\partial z} \right)  \right] \Biggr\rvert_{r = a(z,t)} \nonumber &\\ & = \mathcal{C} \left[\sigma + (\kappa_s - \mu_s) \left(\frac{\left(a u_t\right)'}{a \sqrt{1+a'^2}} + \mathcal{C} u_n\right) \right] \nonumber & \\
& + \frac{2 \mu_s}{1+a'^2} \left[ \frac{ a' u_t + u_n }{a^2 } - \frac{a''}{1+a'^2}\left( u_t' - \frac{a'' u_n}{1+a'^2} \right) \right], \label{eq:bc_normal}
\end{align}
and
\begin{align}
&\frac{\mu}{\sqrt{1+a'^2}} \left[\left(\frac{\partial w}{\partial r} + \frac{\partial u}{\partial z} \right)(1-a'^2) + 2a'\left(\frac{\partial u}{\partial r} - \frac{\partial w}{\partial z} \right) \right] \Biggr\rvert_{r = a(z,t)} = \frac{\partial \sigma}{\partial z}  \nonumber & \\
& +\frac{\partial}{\partial z} \left[(\kappa_s - \mu_s) \left( \frac{\left(a u_t\right)'}{a\sqrt{1+a'^2}} + \mathcal{C} u_n \right) \right] +  \frac{\sqrt{1+a'^2}}{a} \frac{\partial}{\partial{z}} \left[\frac{2 \mu_s a}{1+a'^2} \left( u_t' - \frac{a'' u_n}{1+a'^2} \right) \right]  \nonumber & \\ 
& - \frac{2 \mu_s a'}{\sqrt{1+a'^2}} \left[ \frac{a' u_t + u_n}{a^2} - \frac{a''}{1+a'^2} \left(u_t' - \frac{a'' u_n}{1+a'^2} \right)  \right], \label{eq:bc_tangential}
\end{align}
where $p$ is the pressure, $\mathcal{C} = \bnabla_s \bcdot \boldm{n} = a^{-1}(1+a'^2)^{-1/2} - a'' (1+ a'^2)^{-3/2}$ is twice the mean curvature of the interface and the identity $\bnabla_s \cdot \boldm{u}_s = \bnabla_s \cdot \boldm{u}_t +\mathcal{C} u_n$ has been used. To clarify the derivation of equations~\eqref{eq:bc_normal} and~\eqref{eq:bc_tangential}, several terms of~\eqref{eq:eqsurface} are deduced in detail in appendix~\ref{app:diffgeom}. Note that the left-hand sides of~\eqref{eq:bc_normal} and~\eqref{eq:bc_tangential} are terms arising from the bulk evaluated at $r = a(z,t)$, whereas the terms on their right-hand sides are interface quantities and thus they are previously evaluated at the interface, a fact that must be considered when taking the derivatives of $u_t$ and $u_n$ defined in~\eqref{eq:un_ut}. For instance, the $z$ derivative of the axial velocity at the interface reads, $\partial w(r = a(z,t),z,t)/\partial z = \partial w/ \partial z |_{r = a(z,t)} + a' \partial w/\partial r|_{r = a(z,t)}$.

Equation~\eqref{eq:eqsurface} and its normal and tangential projections in tensor notation are in agreement with those derived by~\cite{Aris1962} and~\cite{Slattery2007}, which correct the typographical errors in~\cite{Scriven60}. However, inasmuch as the boundary conditions~\eqref{eq:bc_normal} and~\eqref{eq:bc_tangential} are necessary for the derivation of the dispersion relation deduced in~\S\ref{sec:results} and the 1D models derived in~\S\ref{sec:longwave}, we present them here in terms of $u$, $w$, $p$ and $a$, and their spatial derivatives. Indeed, to the best of our knowledge,~\eqref{eq:bc_normal} and~\eqref{eq:bc_tangential} have not been reported in the literature in their complete and correct form. Indeed, \cite{Whitaker76} only presents the equations in tensor notation and their linearisation,~\cite{Aris1962} and~\cite{Slattery2007} considered the particular case of a cylinder, $r = R$ and $r = R(t)$, and~\cite[][p. 732]{Slattery2007} presents the projections of the surface equation of motion onto the radial, axial and azimuthal directions in cylindrical coordinates, but not the normal and tangential stress balances.


Furthermore, recent studies in similar axisymmetric configurations have considered surface viscosities in the formulation, e.g.~\cite{Ponce16a, Ponce16b} in the field of liquid bridges, and~\cite{Ponce17} on the breakup of liquid drops. However, in all these cases there are missing terms in both the normal and tangential stress balances. 
The latter mistakes are the reason why the particular case of a cylinder, $r = R(t)$~\citep{Aris1962,Slattery2007} is not recovered, whereas~\eqref{eq:bc_normal} and~\eqref{eq:bc_tangential} correctly reproduce this limit.

\section{Capillary instability of a liquid cylinder with surface viscoelasticity\label{sec:results}}

\subsection{Derivation of the dispersion relation\label{subsec:DR}}
First of all, it is important to emphasise that in the present work we are only concerned with axisymmetric disturbances, since they are the only unstable modes in the case of clean interfaces. Therefore, the axisymmetric continuity, radial momentum and axial momentum equations,
\begin{equation}\label{eq:continuity}
\frac{\partial u}{\partial r} + \frac{\partial w}{\partial z} + \frac{u}{r} = 0,
\end{equation}
\begin{equation}\label{eq:rmomentum}
\rho \left(\frac{\partial u}{\partial t} + u \frac{\partial u}{\partial r} + w \frac{\partial u}{\partial z} \right) = - \frac{\partial p}{\partial r} + \mu \left(\frac{\partial^2 u}{\partial r^2} + \frac{\partial^2 u}{\partial z^2} + \frac{1}{r} \frac{\partial u}{\partial r}  - \frac{u}{r^2} \right),
\end{equation}
\begin{equation}\label{eq:zmomentum}
\rho \left(\frac{\partial w}{\partial t} + u \frac{\partial w}{\partial r} + w \frac{\partial w}{\partial z} \right) = - \frac{\partial p}{\partial z} + \mu \left(\frac{\partial^2 w}{\partial r^2} + \frac{\partial^2 w}{\partial z^2} + \frac{1}{r} \frac{\partial w}{\partial r}  \right),
\end{equation} 
are satisfied by the velocity and pressure fields in the liquid thread.
In addition to the stress balances~\eqref{eq:bc_normal} and~\eqref{eq:bc_tangential}, the kinematic condition has to be satisfied at the free surface $r=a(z,t)$,
\begin{equation}\label{eq:kinematic}
\frac{\partial a}{\partial t} + a' w = u.
\end{equation}
Since we also assume isothermal flow, $\sigma$, $\mu_s$ and $\kappa_s$ are functions only of the surface concentration of surfactant $\Gamma(z,t)$, which satisfies the following transport equation at $r = a(z,t)$:
\begin{equation}
\frac{\partial \Gamma}{\partial t} + w \frac{\partial \Gamma}{\partial z} + \frac{\Gamma}{a \sqrt{1+a'^2}} \frac{\partial (a u_t)}{\partial z} + \mathcal{C} \Gamma  u_n = 0.\label{eq:surftransport}
\end{equation}
Note that the time derivative in~\eqref{eq:surftransport} is referred to a frame of reference fixed in space, i.e. the laboratory frame -- for details on the subtleties of the time derivative of a surface quantity, see~\citet{Stone90, Wong1996, Pereira2008}. Furthermore, the surface diffusion has been neglected due to the large typical values of the associated P\'eclet number~\citep{Timmermans02}. Indeed, the typical values of surfactant surface diffusivities $D_s$ are in the range of $10^{-10}-10^{-9}$ m$^2$ s$^{-1}$~\citep{Tricot1997,Valkovska2000, Liao2006}, which leads to surface P\'eclet numbers, defined as $Pe_s = R^2/(D_s t_c) = [\sigma_0 R/(\rho D_s^2)]^{1/2}$, of order $Pe_s\sim 10^{6}-10^{5}$ for a water thread of 1~mm radius and taking the capillary time as characteristic time scale, $t_c = (\rho R^3/\sigma_0)^{1/2}$.

All the flow variables are assumed to be slightly perturbed around a stationary and uniform state in which the liquid thread is considered infinitely long with an initial cylindrical shape of radius $R$ (see figure~\ref{fig:figure1}),
\begin{align}
(u,w,p,\sigma,\Gamma,a,\kappa_s,\mu_s) = (0,0,p_a+\sigma_0/R,\sigma_0, \Gamma_0, R,\kappa_{s0},\mu_{s0})  \nonumber \\ +\epsilon \, (\delta u,\delta w,\delta p,\delta \sigma, \delta \Gamma, \delta a, \delta \kappa_s, \delta \mu_s),
\end{align}
where $\epsilon \ll 1$ and $\sigma_0$, $\kappa_{s0}$ and $\mu_{s0}$ are the uniform values of surface tension, dilatational and shear surface viscosity respectively, associated with an initially uniform superficial surfactant concentration, $\Gamma_0$. Note that the reference frame has been selected such that the unperturbed thread is at rest. The linearised continuity, radial momentum and axial momentum equations read, respectively,
\begin{equation}
\frac{\partial (\delta u)}{\partial r} + \frac{\partial ( \delta w)}{\partial z} + \frac{\delta u}{r} = 0, \label{eq:lin_cont}
\end{equation}
\begin{equation}
\rho \frac{\partial ( \delta u)}{\partial t}  = - \frac{\partial ( \delta p)}{\partial r} + \mu \left(\frac{\partial^2 (\delta u)}{\partial r^2} + \frac{\partial^2 ( \delta u)}{\partial z^2} + \frac{1}{r} \frac{\partial (\delta u)}{\partial r}  - \frac{\delta u}{r^2} \right), \label{eq:lin_momr}
\end{equation}
\begin{equation}
\rho \frac{\partial (\delta w)}{\partial t} = - \frac{\partial (\delta p)}{\partial z} + \mu \left(\frac{\partial^2 (\delta w)}{\partial r^2} + \frac{\partial^2 (\delta w)}{\partial z^2} + \frac{1}{r} \frac{\partial (\delta w)}{\partial r}  \right), \label{eq:lin_momz}
\end{equation}
with a harmonic pressure disturbance field,
\begin{equation}
\nabla^2(\delta p) = \frac{\partial^2 (\delta p)}{\partial r^2} + \frac{1}{r} \frac{\partial (\delta p)}{\partial r} + \frac{\partial^2 (\delta p)}{\partial z^2} = 0.
\end{equation}
Finally, the linearised boundary conditions at the free surface, $r=R$, are 
\begin{eqnarray}
  &&   \delta p - 2 \mu \frac{\partial (\delta u )}{\partial r} = \frac{\delta \sigma}{R} - \sigma_0 \left(\frac{\partial^2 (\delta a)}{\partial z^2} + \frac{\delta a}{R^2} \right) + \frac{\kappa_{s0} - \mu_{s0}}{R} \frac{\partial ( \delta w)}{\partial z} +  \frac{\kappa_{s0} + \mu_{s0}}{R^2} \delta u, \label{eq:bc1} \\
  &&    \mu \left(\frac{\partial (\delta w)}{\partial r} + \frac{\partial (\delta u)}{\partial z}  \right) = \frac{\partial (\delta \sigma)}{\partial z} + (\kappa_{s0} + \mu_{s0})\frac{\partial^2 (\delta w)}{\partial z^2} + \frac{\kappa_{s0} - \mu_{s0}}{R} \frac{\partial(\delta u)}{\partial z}, \label{eq:bc2} \\ 
  &&    \frac{\partial (\delta a)}{\partial t} = \delta u, \label{eq:bc3} \\
  &&    \frac{\partial (\delta \Gamma)}{\partial t} + \Gamma_0 \left(\frac{\partial (\delta w)}{\partial z} + \frac{\delta u}{R} \right) = 0, \label{eq:bc4} \\
  &&    \delta \sigma = -\frac{E}{\Gamma_0} \delta \Gamma, \label{eq:bc5}
\end{eqnarray}
representing, respectively, the normal and tangential stress conditions, the kinematic condition, the surfactant transport equation and the constitutive equation for the dependence of surface tension on surfactant concentration, $\sigma(\Gamma)$, where $E = - \Gamma_0 \left.(\partial \sigma/ \partial \Gamma)\right|_{\Gamma_0}$ is the Gibbs elasticity. The linearised versions of the stress boundary conditions taking into account the effect of surface viscosities, i.e.~\eqref{eq:bc1} and~\eqref{eq:bc2}, are identical to those of~\cite{Whitaker76}. To derive a dispersion relation $D(k,\omega)=0$ between the axial wavenumber $k$ and the temporal growth rate $\omega$, the disturbances are decomposed as normal modes,
\begin{equation}
(\delta u,\delta w,\delta p,\delta \sigma, \delta \Gamma, \delta a) = (\hat{u}(r),\hat{w}(r),\hat{p}(r),\hat{\sigma},\hat{\Gamma},\hat{a})\exp(ikz + \omega t).
\end{equation}
where the pressure amplitude $\hat{p}$ satisfies the modified Bessel equation,
\begin{equation}
\frac{\text{d}^2 \hat{p}}{\text{d} r^2} + \frac{1}{r}\frac{\text{d} \hat{p}}{\text{d} r} - k^2 \hat{p} = 0,
\end{equation}
with regular solution
\begin{equation}
\hat{p}(r) = A \, I_0(kr),
\end{equation}
where $A$ is an integration constant and $I_n(kr)$ is the $n$th-order modified Bessel function of the first kind. The linearised system of~\eqref{eq:lin_cont}--\eqref{eq:lin_momz} and~\eqref{eq:bc3}--\eqref{eq:bc5} can be straightforwardly solved for $\hat{u}(r)$, $\hat{w}(r)$, $\hat{a}$, $\hat{\Gamma}$ and $\hat{\sigma}$, to give
\begin{equation}
\hat{u}(r) = B I_1(\tilde{k}r) - A \frac{k}{\rho \omega} I_1(k r),
\end{equation}
\begin{equation}
\hat{w}(r) = B \frac{i \tilde{k}}{k} I_0(\tilde{k}r) - A \frac{i k}{\rho \omega} I_0(k r),
\end{equation}
\begin{equation}
\hat{a} = \frac{\hat{u}(R)}{\omega} = B \frac{I_1(\tilde{k}R)}{\omega} - A \frac{k}{\rho \omega^2}I_1(k R),
\end{equation}
\begin{equation}
\hat{\Gamma} = B \frac{\Gamma_0}{\omega} \left[\tilde{k} I_0(\tilde{k}R) - \frac{I_1(\tilde{k}R)}{R}\right] - A \frac{k \Gamma_0}{\rho \omega^2} \left[k I_0(k R)-\frac{I_1(k R)}{R} \right],
\end{equation}
\begin{equation}
\hat{\sigma} = -B \frac{E}{\omega} \left[\tilde{k} I_0(\tilde{k}R) - \frac{I_1(\tilde{k}R)}{R}\right] + A \frac{k E}{\rho \omega^2} \left[k I_0(k R)-\frac{I_1(k R)}{R} \right],
\end{equation}
where $B$ is another integration constant and $\tilde{k}^2 = k^2 + \rho \omega/\mu$. 
Finally, the normal and tangential stress conditions given by equations~\eqref{eq:bc1} and~\eqref{eq:bc2} provide, respectively,
\begin{equation}
\hat{p}(R) = 2 \mu \frac{\text{d} \hat{u}}{\text{d} r}(R) + \frac{\hat{\sigma}}{R} - \frac{\sigma_0}{\omega R^2}(1-k^2 R^2) \hat{u}(R) + \frac{\kappa_{s0}-\mu_{s0}}{R}ik\hat{w}(R) + \frac{\kappa_{s0} + \mu_{s0}}{R^2} \hat{u}(R), \label{eq:bc_modes_normal}
\end{equation}
\begin{equation}
\mu \left[\frac{\text{d}\hat{w}}{\text{d} r}(R) + i k \hat{u}(R) \right] = ik \hat{\sigma} - k^2(\kappa_{s0}+\mu_{s0}) \hat{w}(R) + \frac{i k (\kappa_{s0} - \mu_{s0})}{R} \hat{u}(R). \label{eq:bc_modes_tangential}
\end{equation}
Once the normal-mode variables evaluated at $r=R$ are substituted in~\eqref{eq:bc_modes_normal} and~\eqref{eq:bc_modes_tangential}, these provide a homogeneous linear system for $A$ and $B$, namely $\mathsfbi{M} \bcdot \boldsymbol{\phi} = 0$, where $\boldsymbol{\phi} = (A,B)^T$. The four entries of $\mathsfbi{M}$ can be found in appendix~\ref{app:matrix_entries}. Non-trivial solutions require that $\det(\mathsfbi{M}) = 0$, finally leading to the desired dispersion relation,
\begin{align}
& \Reycap \omega^2 F(k) - k^2(1-k^2)+ \beta k^2[1+F(k)(F(\tilde{k})-2)] +\frac{k^4}{\Reycap}\left[4 - \frac{\beta}{\omega} \left(2 - \frac{1-k^2}{\omega} + 4 \Bou_{\mu} \right)\right. \nonumber & \\
& \left. + 6\Bou_{\mu} + \frac{1-k^2}{\omega}(\Bou_{\mu} + \Bou_{\kappa}) - 2 \Bou_{\kappa} (1 + 2\Bou_{\mu})\right][F(k)-F(\tilde{k})] +\omega k^2 \left[2(\Bou_{\mu} - \Bou_{\kappa})F(k) \right. \nonumber & \\
& \left. + (\Bou_{\mu} + \Bou_{\kappa})(F(k)F(\tilde{k}) + 1) + 2(2F(k) -1)\right]= 0, \label{eq:dimless_DR}
\end{align}
where $\tilde{k}^2 = k^2 + \Reycap \omega$ and $F(x) = x I_0(x)/I_1(x)$. Equation~\eqref{eq:dimless_DR} has been non-dimensionalised taking $R$ as characteristic length and the viscocapillary time, $\mu R/ \sigma_0$ as characteristic time, where, for the sake of clarity, we have kept the same notation $\omega$ and $k$ for the dimensionless growth rate and wavenumber, respectively. According to~\eqref{eq:dimless_DR}, the dispersion relation depends on four dimensionless parameters: the capillary Reynolds number, $\Reycap = \text{\emph{Oh}}^{-2} = \rho \sigma_0 R/\mu^2$, where $\Oh$ is the Ohnesorge number; the elasticity parameter $\beta = E/\sigma_0$; and the shear and dilatational Boussinesq numbers, $\Bou_{\mu} = \mu_{s0}/(\mu R)$ and $\Bou_{\kappa} = \kappa_{s0}/(\mu R)$, respectively. In presenting the results below, the parameter $\Bou_{\kappa}$ will be substituted by the ratio of surface shear to dilatational viscosity, $\mu_{s0}/\kappa_{s0}=\Bou_{\mu}/\Bou_{\kappa}$. Note that the dispersion relation~\eqref{eq:dimless_DR} reduces to that of~\citet{Timmermans02} in the limit $\Bou_{\mu}\to 0$ and $\Bou_{\kappa}\to 0$, and to the Rayleigh--Chandrasekhar dispersion relation when, in addition, $\beta \to 0$~\citep{Rayleigh4,Chandrasekhar}. However, there are several differences between~\eqref{eq:dimless_DR} and the dispersion relation deduced by~\citet{Whitaker76} in the insoluble limit due to several mistakes made in the latter work~\citep{Hansen99,Timmermans02}.

\subsection{Temporal stability analysis\label{subsec:temporal}}

\begin{figure}
    \centering
    \includegraphics[width=0.9\textwidth]{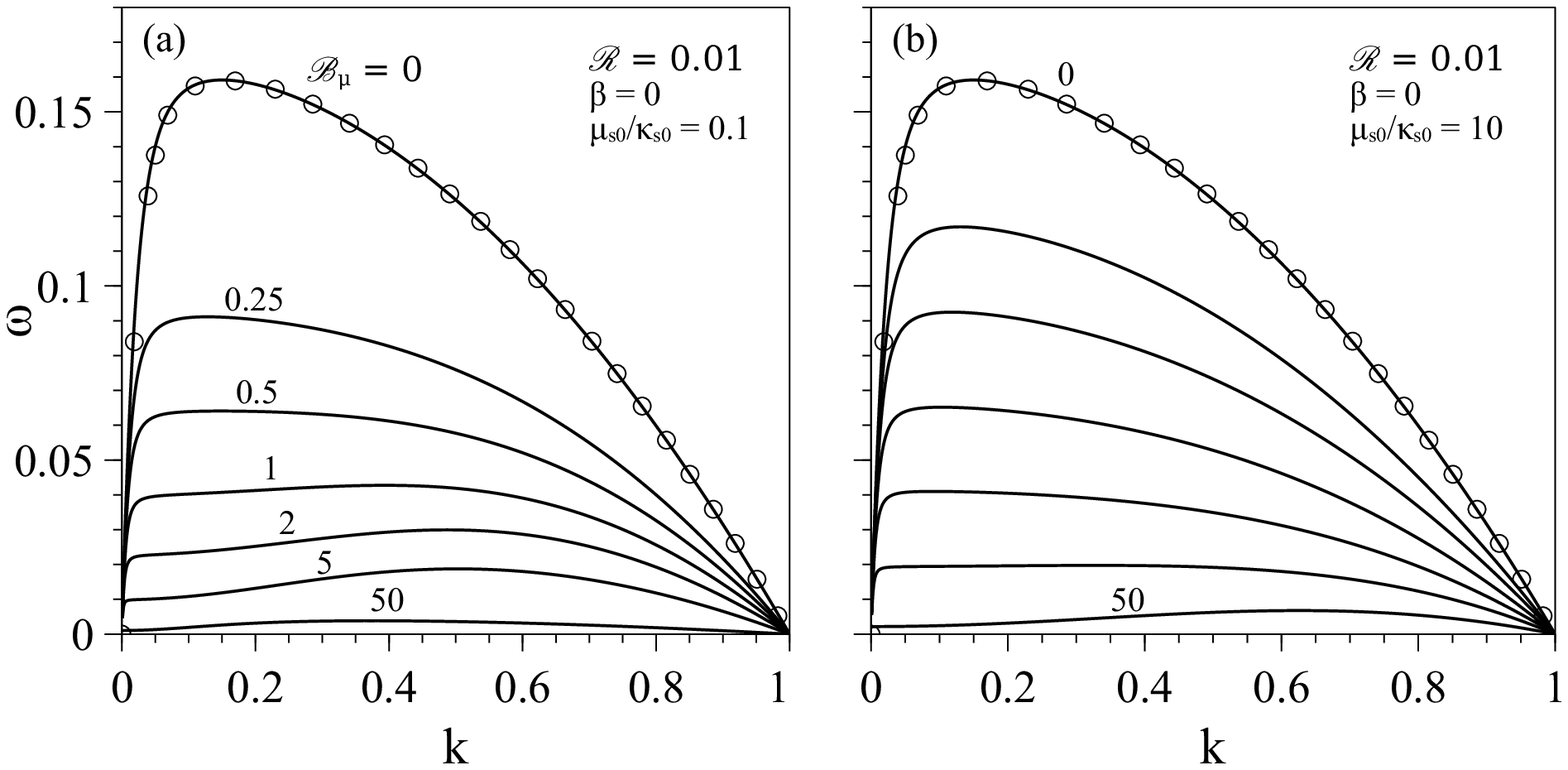}
    \includegraphics[width=0.9\textwidth]{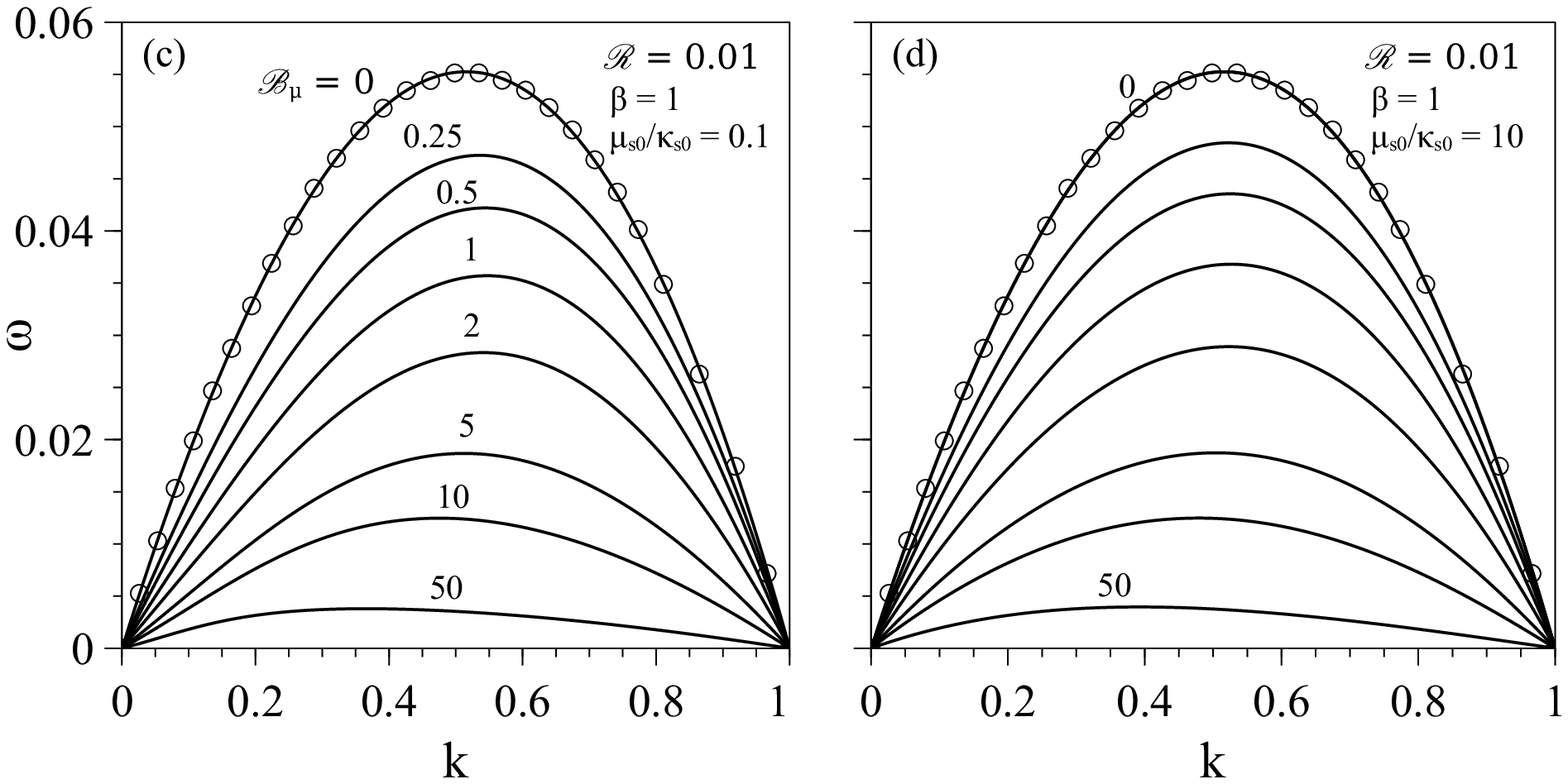}
    \caption{Growth rate $\omega$ as a function of wavenumber $k$ for $\Reycap=0.01$ and $\beta=0$ ($a$,$b$), $\beta = 1$ ($c$,$d$), $\mu_{s0}/\kappa_{s0}=0.1$ ($a$,$c$) and $\mu_{s0}/\kappa_{s0}=10$ ($b$,$d$). The values of $\Bou_{\mu}$ are indicated near each curve. The symbols are extracted from figure~4 of~\citet{Timmermans02}.\label{fig:figure2}}
\end{figure}

\begin{figure}
    \centering
    \includegraphics[width=0.9\textwidth]{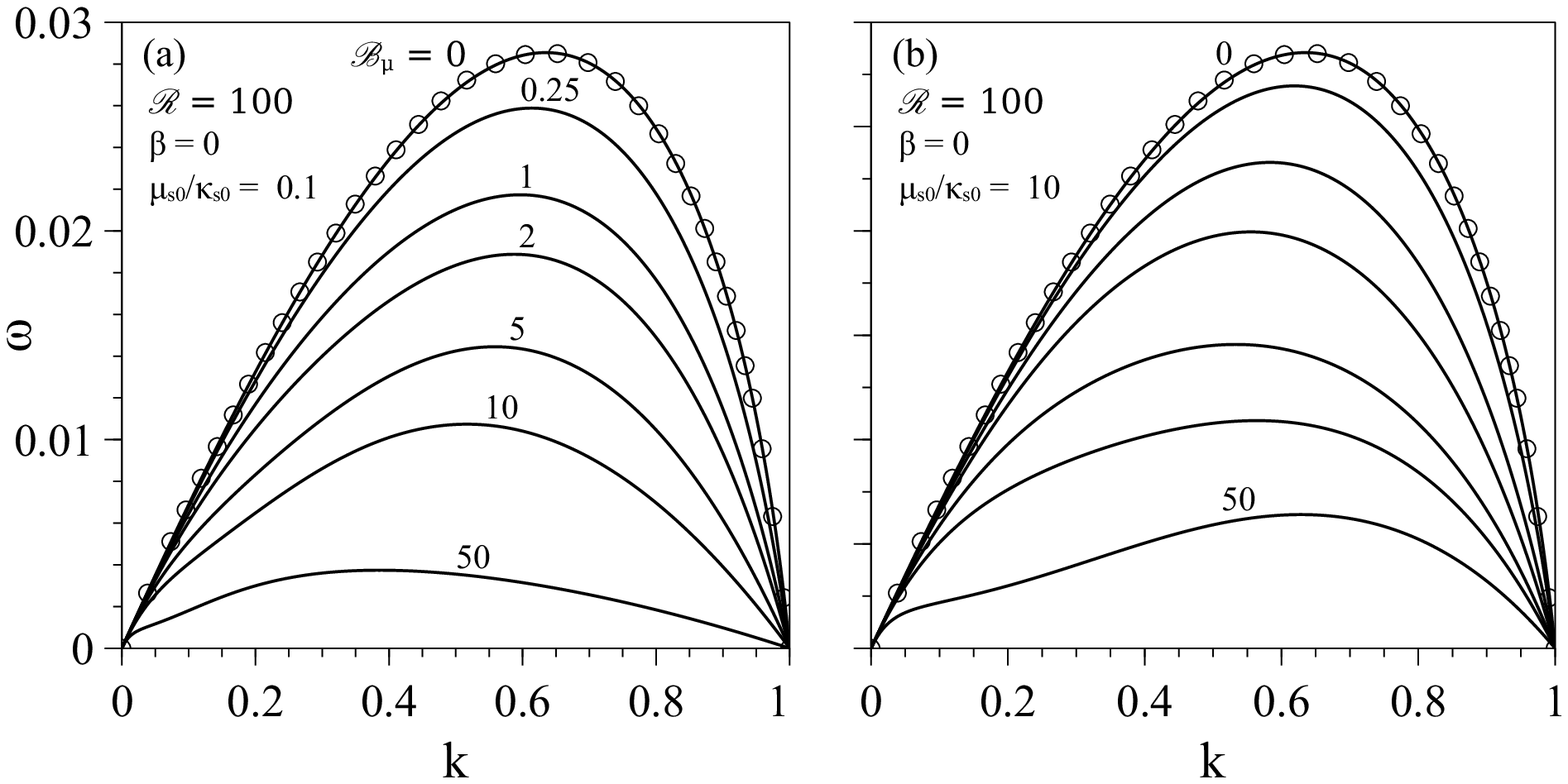}
    \includegraphics[width=0.9\textwidth]{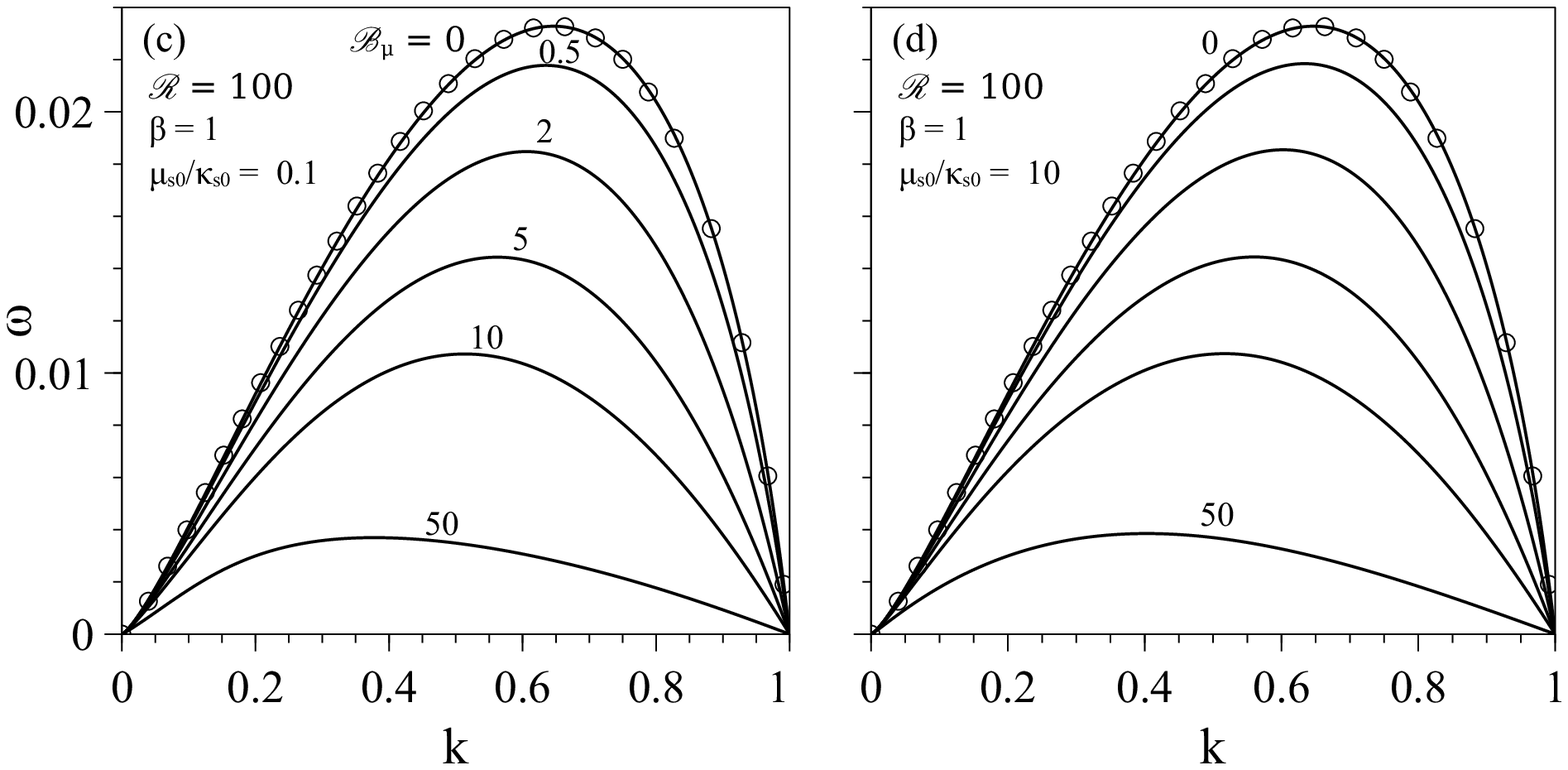}
    \caption{Same as figure~\ref{fig:figure2} for $\Reycap=100$.\label{fig:figure3}}
\end{figure}

To illustrate the effect of surface viscosities on the growth rate of infinitesimal disturbances on the liquid cylinder, figures~\ref{fig:figure2} and~\ref{fig:figure3} show the amplification function $\omega(k)$ for $\Reycap=0.01$ and $\Reycap=100$, respectively, with the value of $\Bou_{\mu}$ indicated near each curve. In figures~\ref{fig:figure2} and~\ref{fig:figure3}, panels ($a$,$b$) and ($c$,$d$) correspond to values of the elasticity parameter $\beta=0$ and $\beta=1$, while panels ($a$,$c$) and ($b$,$d$) show the results obtained with values of the surface viscosity ratio $\mu_{s0}/\kappa_{s0}=0.1$ and $10$, respectively. Our results were validated by comparing them with four amplification curves extracted from figure~4 of~\citet{Timmermans02}, represented by symbols, for the particular case $\Bou_{\mu}=0$, finding perfect agreement.

As revealed by figures~\ref{fig:figure2} and~\ref{fig:figure3}, surface viscosity stabilises the liquid cylinder, since the values of $\omega$ decrease monotonically for all values of $k$ as $\Bou_{\mu}$ increases. The stabilising effect is more pronounced for small values of $\Reycap$ and $\beta$. Indeed, in the particular case with $\Reycap=0.01$ and $\beta=0$, figure~\ref{fig:figure2}($a$,$b$) shows that, for a value of $\Bou_{\mu}=1$, the maximum growth rate decreases by factors of approximately 3.8 and 2.4, respectively, compared with the case with $\Bou_{\mu}=0$. These factors are reduced to approximately 1.5 in figures~\ref{fig:figure2}(c) and~\ref{fig:figure2}(d), showing that an increase in the surface elasticity decreases the relative importance of surface viscosity. 

It is important to point out that increasing the value of $\beta$ also has a stabilising effect, as studied in detail by~\citet{Timmermans02}, and also clearly seen in figures~\ref{fig:figure2} and~\ref{fig:figure3}. However, in contrast with the fact that a finite value of $\omega$ is achieved when $\beta\to\infty$, our results reveal that $\omega\to 0$ when $\Bou_{\mu}\to \infty$, a behaviour similar to that observed for $\Reycap\to 0$, and clearly due to the dissipative nature of the surface shear viscosity (see \S\ref{subsubsec:Bmugg}). In particular, the effect of $\Bou_{\mu}$ is important for both for low- and high-viscosity threads although, as happens with the effect of $\beta$~\citep{Timmermans02}, the effect is more pronounced at low capillary Reynolds numbers, as can be deduced by comparing the results of figure~\ref{fig:figure2} for $\Reycap=0.01$ with those of figure~\ref{fig:figure3} for $\Reycap=100$. Indeed, the reduction factors in the maximum growth rate between the cases $\Bou_{\mu}=0$ and $\Bou_{\mu}=1$ are substantially smaller for $\Reycap=100$ compared with those stated before for $\Reycap=0.01$. In particular, these factors are approximately 1.3 and 1.2 in figures~\ref{fig:figure3}(a) and~\ref{fig:figure3}(b), respectively. The latter limit of dominant inertia, $\Reycap\gg 1$, was studied by~\citet{Whitaker76}, whose conclusions about the effect of the surface viscosities on $\omega$ were correct and coincide with the ones presented herein.

By comparing the left and right columns of figures~\ref{fig:figure2} and~\ref{fig:figure3} it is deduced that $\omega$ is only slightly affected by $\mu_{s0}/\kappa_{s0}$, with the general trend that an increase in $\Bou_{\kappa}$ decreases the growth rate, as expected. In particular, for $\beta \lesssim 1$ the maximum growth rate, $\omega_m$, is smaller when $\Bou_{\kappa}$ is greater than $\Bou_{\mu}$, as depicted by figures~\ref{fig:figure2}($a$,$b$) and~\ref{fig:figure3}($a$,$b$), this phenomenon being more noticeable for $\Reycap \ll 1$. This result is expected, since for a fixed $\Bou_{\mu}$ the overall surface viscosity increases when $\mu_{s0}/\kappa_{s0}$ decreases, as already noticed by~\citet{Whitaker76} in the limit $\Reycap\gg 1$. Figures~\ref{fig:figure2} and~\ref{fig:figure3} also reveal that the wavenumber of maximum amplification, $k_m$, depends non-monotonically on $\Bou_{\mu}$ for a fixed value of the surface viscosity ratio, most notably at low capillary Reynolds numbers and large values of $\mu_{s0}/\kappa_{s0}$.

\subsubsection{\label{subsubsec:Bmugg}The limit $\Bou_{\mu} \to \infty$ for finite $\beta$ and $\Bou_{\mu}/\Bou_{\kappa}$.}

\begin{figure}
    \centering
    \includegraphics[width=0.9\textwidth]{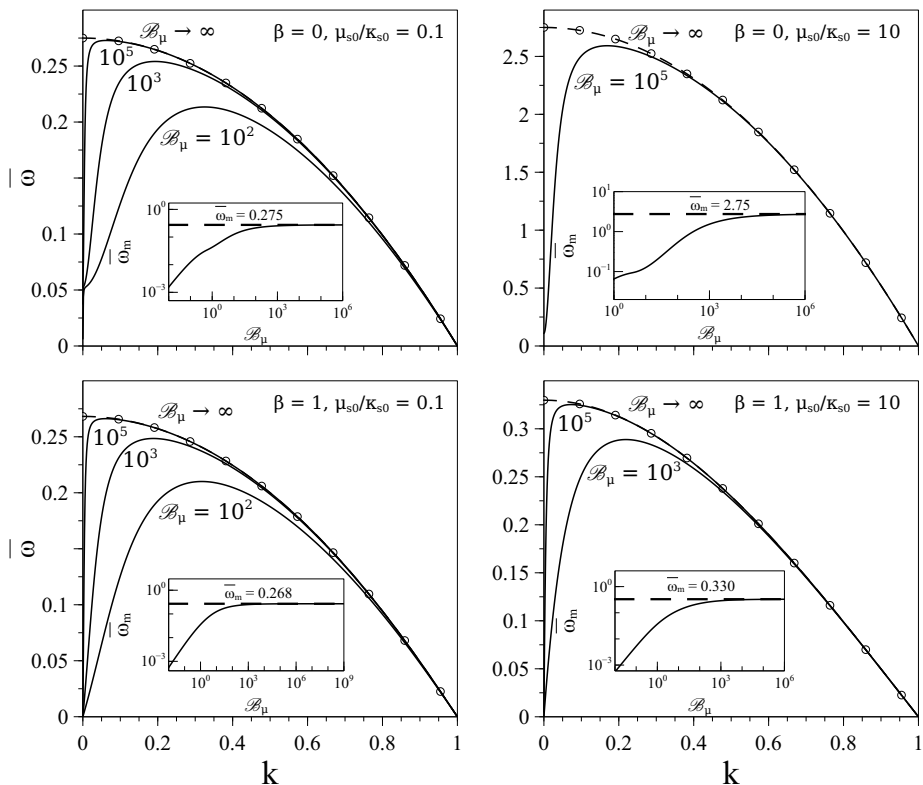}
    \caption{Rescaled growth rate $\bar{\omega}=\omega\,\Bou_{\mu}$ as a function of $k$, computed with~\eqref{eq:dimless_DR}, for $\beta=0$ ($a$,$b$), $\beta = 1$ ($c$,$d$), $\mu_{s0}/\kappa_{s0}=0.1$ ($a$,$c$) and $\mu_{s0}/\kappa_{s0}=10$ ($b$,$d$). The dashed line with open circles is the amplification curve given by~\eqref{eq:Bmugg} in the limit $\Bou_{\mu} \gg 1$. The insets show the dependence of the rescaled maximum growth rate, $\bar{\omega}_{m}$, on $\Bou_{\mu}$, while the dashed line is the horizontal asymptote predicted by~\eqref{eq:Bmugg} evaluated at $k=0$. \label{fig:figure4}}
\end{figure}

As depicted in figures~\ref{fig:figure2} and~\ref{fig:figure3}, in the limit $\Bou_{\mu} \to \infty$ the temporal growth rate $\omega\to 0$ for every value of $k$ within the unstable range, $0\leq k\leq 1$. This result indicates that the viscocapillary time, $\mu R/ \sigma_0$, is not the characteristic time of the Plateau-Rayleigh instability when $\Bou_{\mu}\gg 1$. Instead, the appropriate time scale in this limit is the surface-shear viscocapillary time, $\mu_{s0}/\sigma_0$. If the latter is used to define a new dimensionless growth rate, $\bar{\omega} = \omega \, \Bou_{\mu}$, the dispersion relation~\eqref{eq:dimless_DR} reduces, in the limit $\Bou_{\mu} \gg 1$, to
\begin{equation}\label{eq:Bmugg}
\bar{\omega} = \left(1 + \frac{\Bou_{\mu}}{\Bou_{\kappa}}\right)\frac{1-k^2}{8} - \frac{\beta}{2} \frac{\Bou_{\mu}}{\Bou_{\kappa}} + \sqrt{ \beta \frac{\Bou_{\mu}}{\Bou_{\kappa}} \frac{1-k^2}{4} + \left[ \frac{\beta}{2} \frac{\Bou_{\mu}}{\Bou_{\kappa}} - \left(1 + \frac{\Bou_{\mu}}{\Bou_{\kappa}} \right) \frac{1-k^2}{8}  \right]^2 }+ O(\Bou_{\mu}^{-1})
\end{equation}
where only leading-order terms have been retained. Note that~\eqref{eq:Bmugg} depends on $\beta$ and $\Bou_{\mu}/\Bou_{\kappa}=\mu_{s0}/\kappa_{s0}$, but it is independent of the capillary Reynolds number, $\Reycap$. Consequently, neither the inertia of the liquid nor the bulk viscous stress play any role in the limit $\Bou_{\mu} \to \infty$, since the flow is dominated by surface stresses. It is also deduced from~\eqref{eq:Bmugg} that the amplification curve $\bar{\omega}(k)$ reaches its maximum at $k=0$, similarly to what happens in the Stokes limit $\Reycap \to 0$. This fact is illustrated in figure~\ref{fig:figure4}, which shows the amplification curves obtained from the exact dispersion relation provided in~\eqref{eq:dimless_DR} for several values of $\Bou_{\mu}$ and two different values of $\mu_{s0}/\kappa_{s0}$ and of $\beta$, and where the dashed line with open circles represents~\eqref{eq:Bmugg}. The insets show that the rescaled maximum growth rates, $\bar{\omega}_{m}$ (solid lines), tend to a finite value when $\Bou_{\mu} \to \infty$, which can be easily computed as a function of $\beta$ and $\mu_{s0}/\kappa_{s0}$ evaluating~\eqref{eq:Bmugg} at $k=0$ (dashed lines).

\subsubsection{\label{subsubsec:omegam_km} Analysis of the maximum growth rate $\omega_m$ and its associated wavenumber $k_m$}

\begin{figure}
 \includegraphics[width=\textwidth]{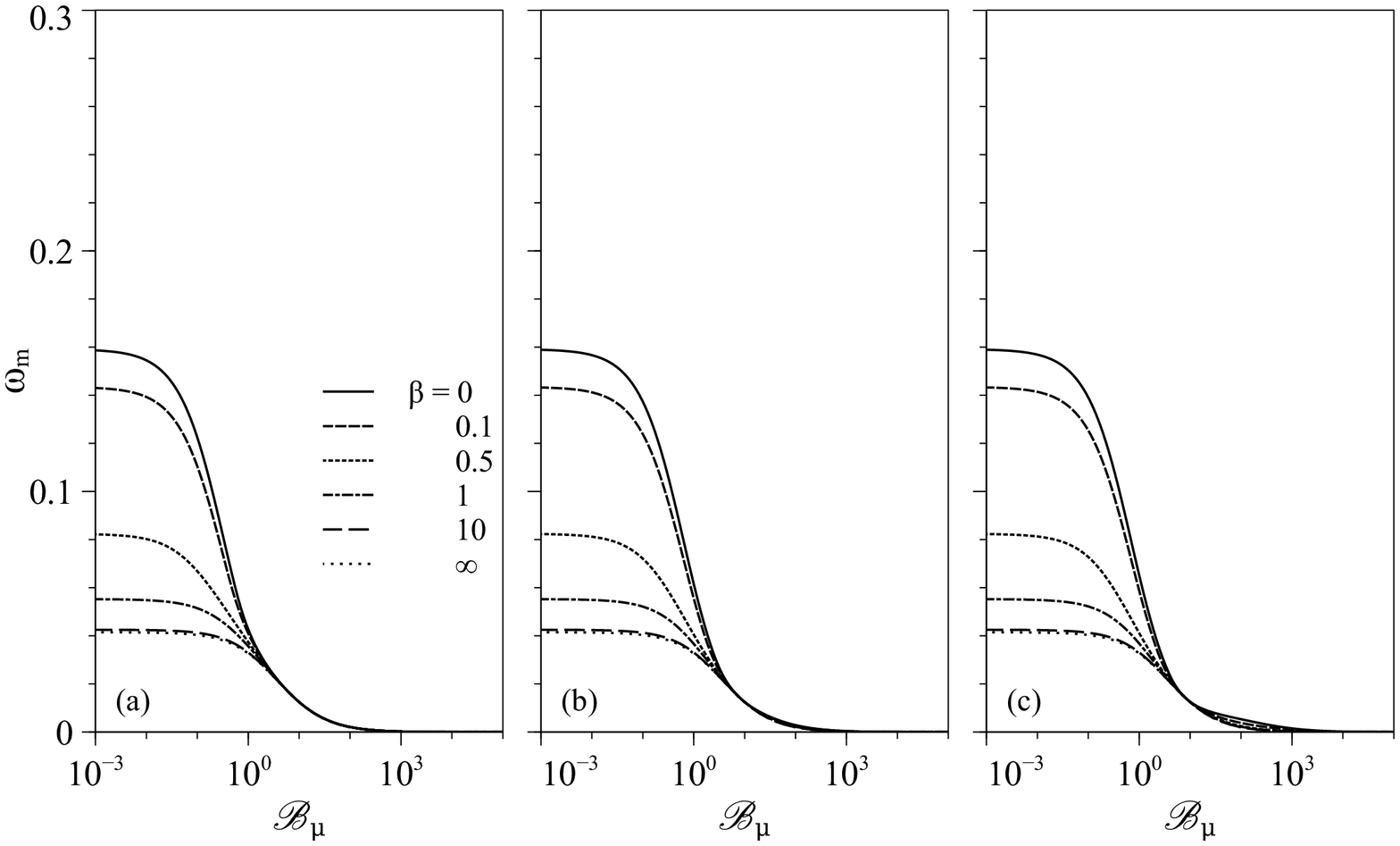}%
 \begin{picture}(0,0)
 \put(-115,137.5){\includegraphics[height=2.8cm]{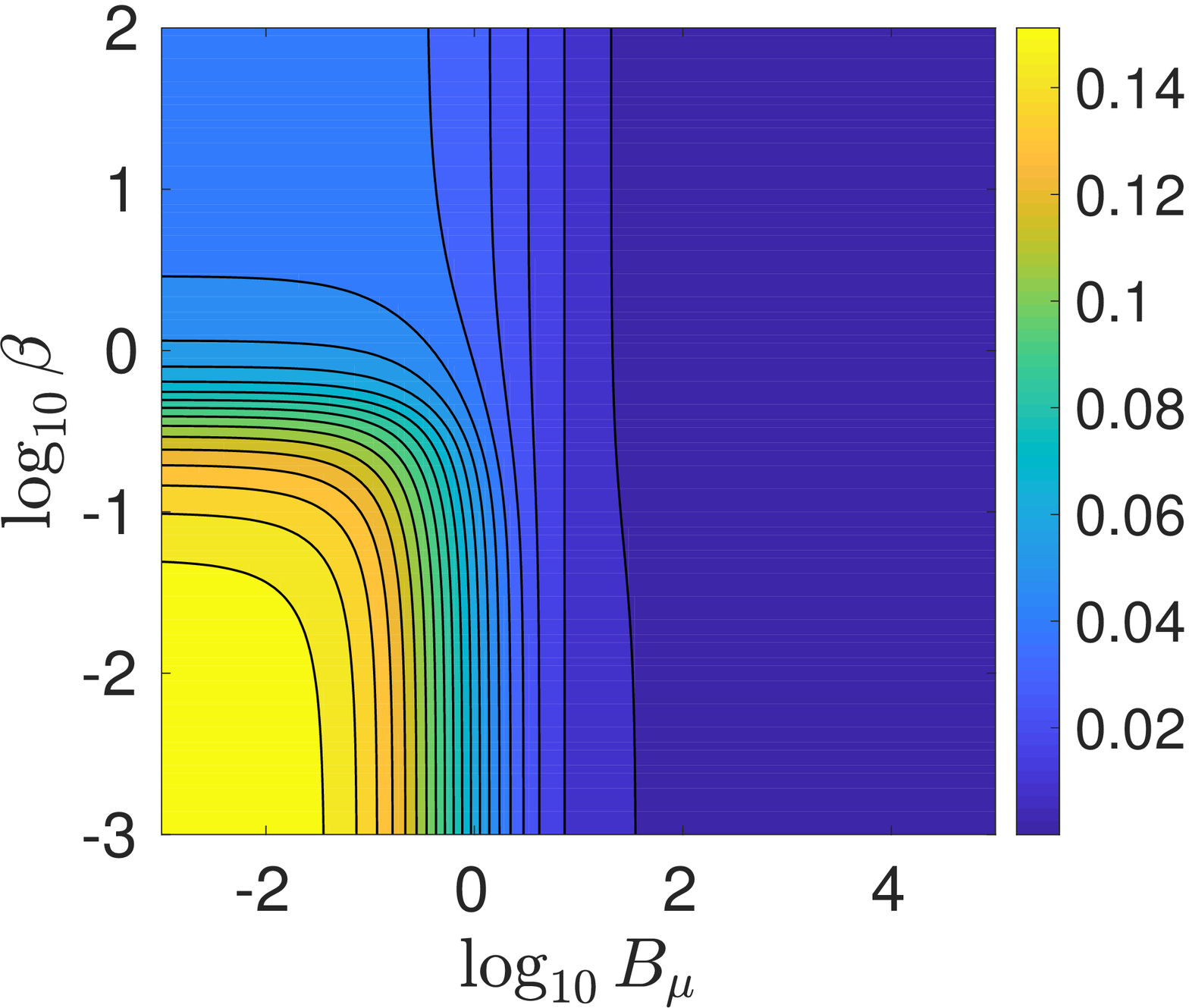}}
 \end{picture}
 \begin{picture}(0,0)
 \put(150,147.5){\includegraphics[height=2.8cm]{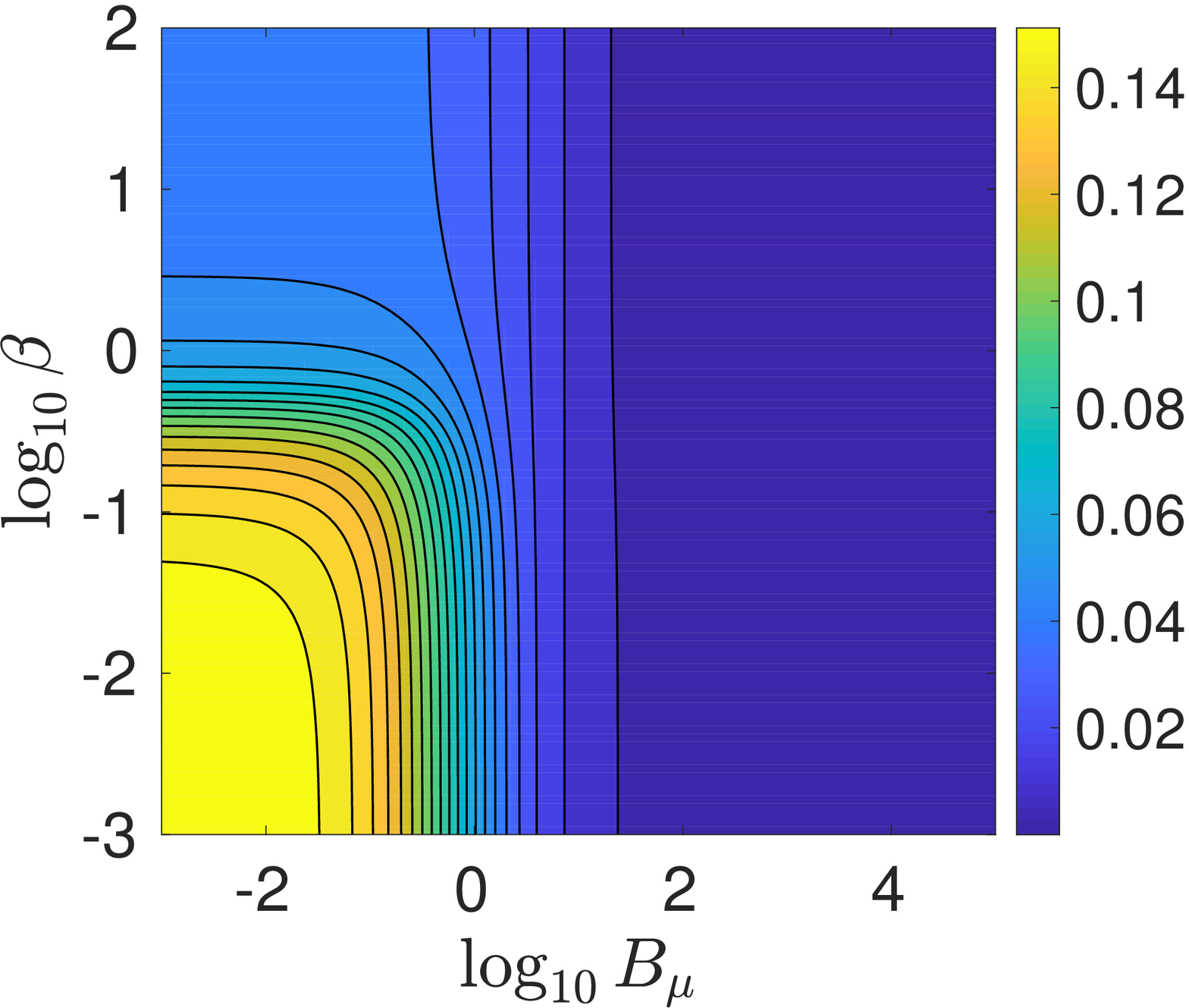}}
 \end{picture}
 \begin{picture}(0,0)
 \put(30,147.5){\includegraphics[height=2.8cm]{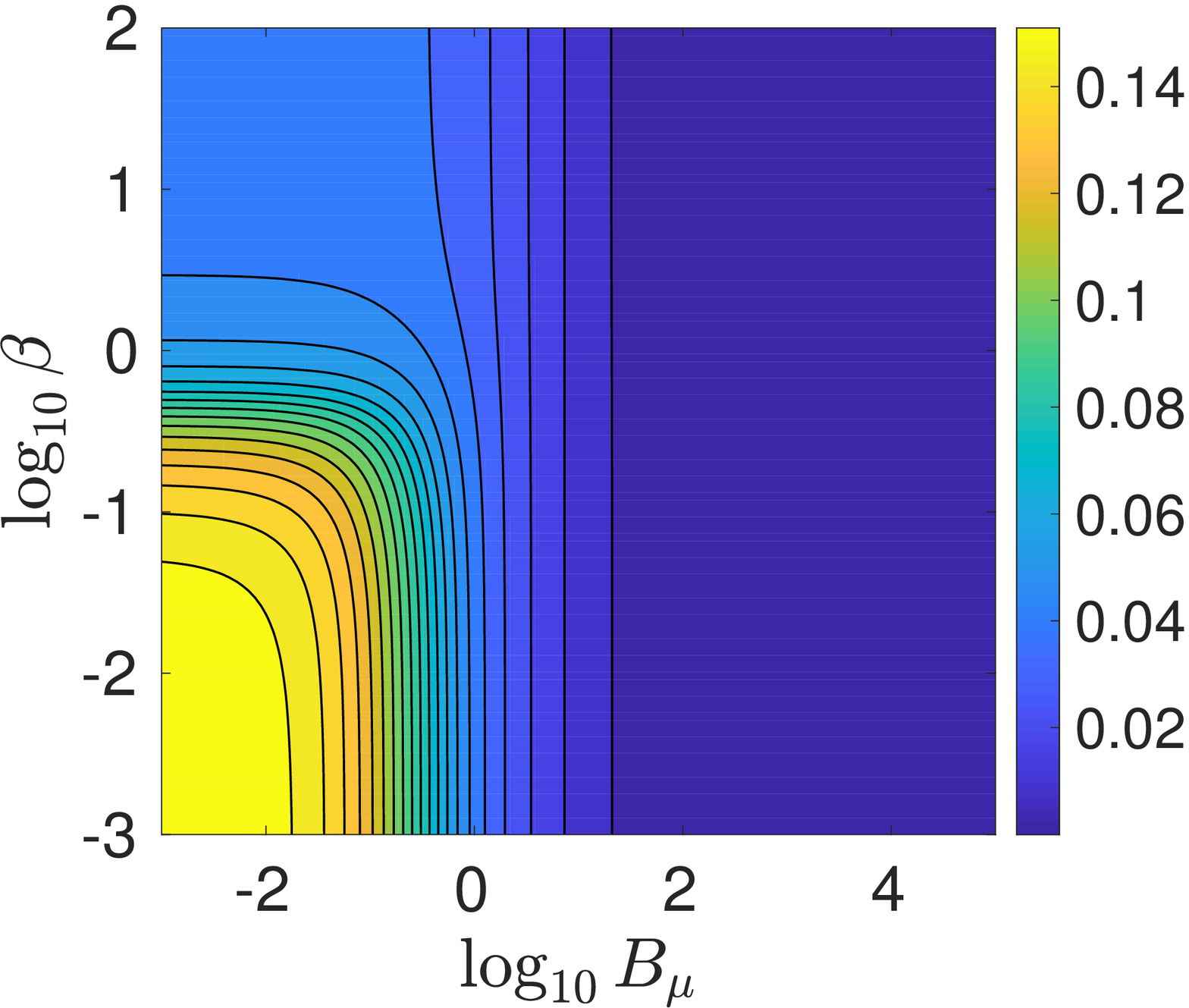}}
 \end{picture}\\
 \includegraphics[width=\textwidth]{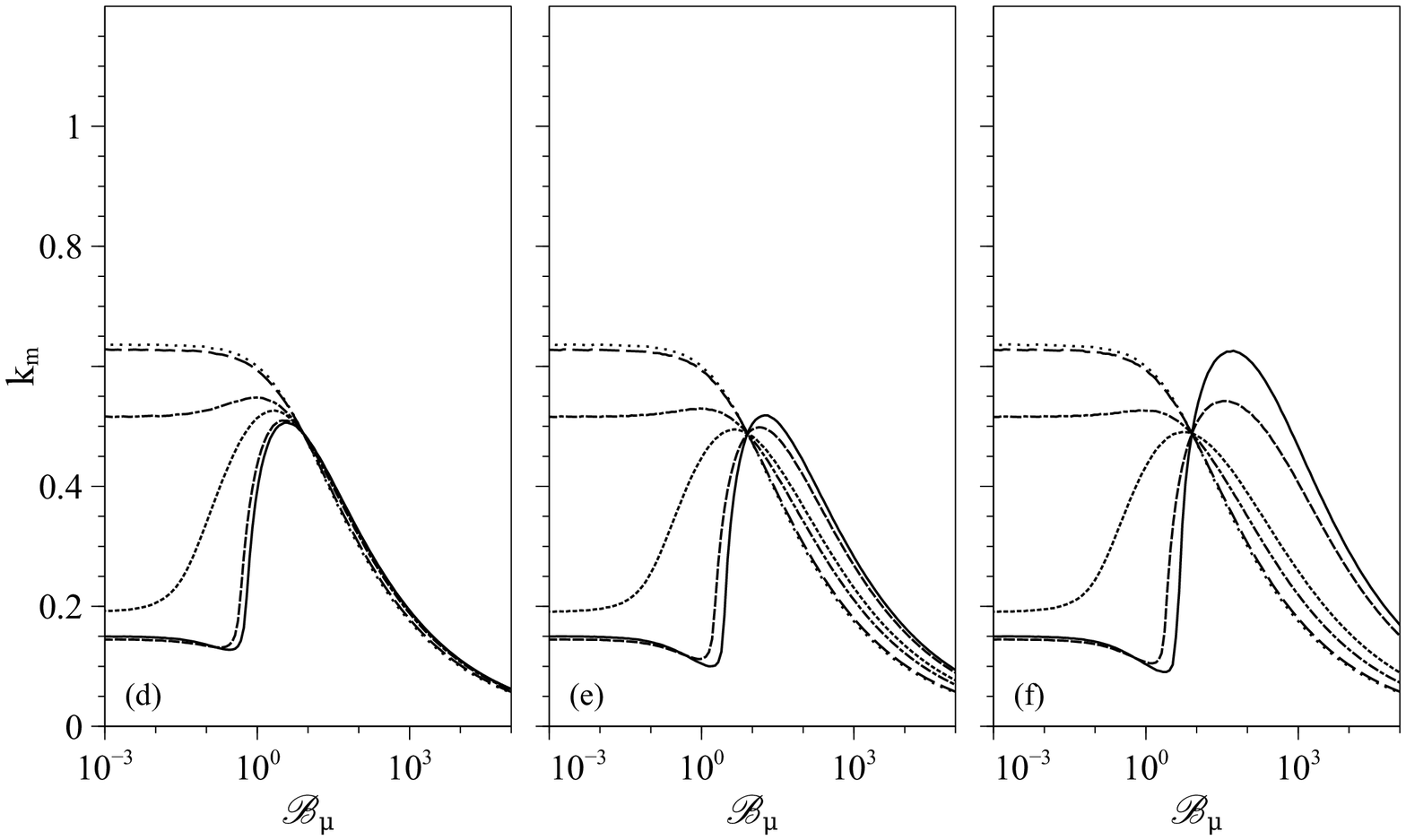}%
 \begin{picture}(0,0)
 \put(-115,137.5){\includegraphics[height=2.8cm]{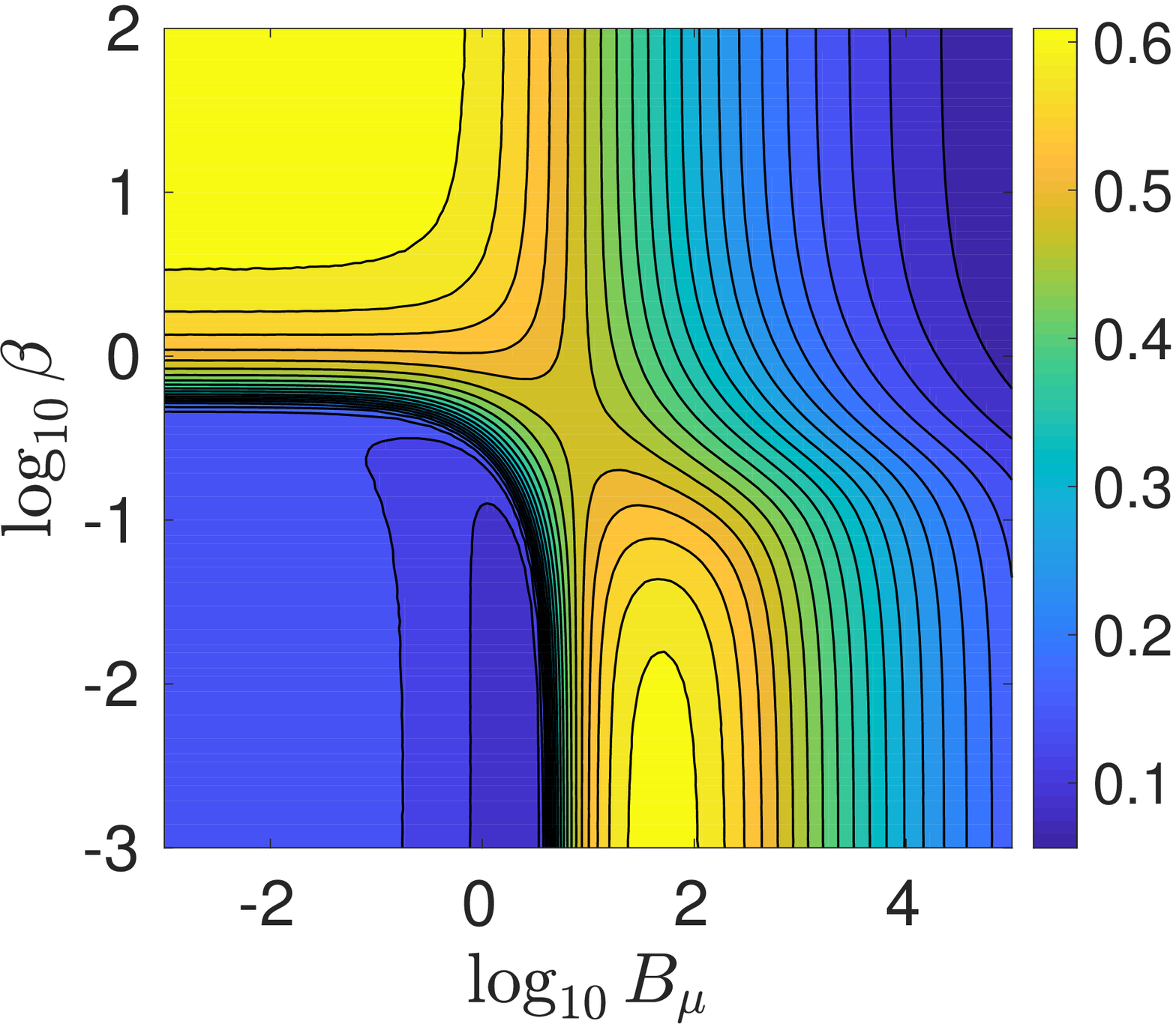}}
 \end{picture}
 \begin{picture}(0,0)
 \put(150,147.5){\includegraphics[height=2.8cm]{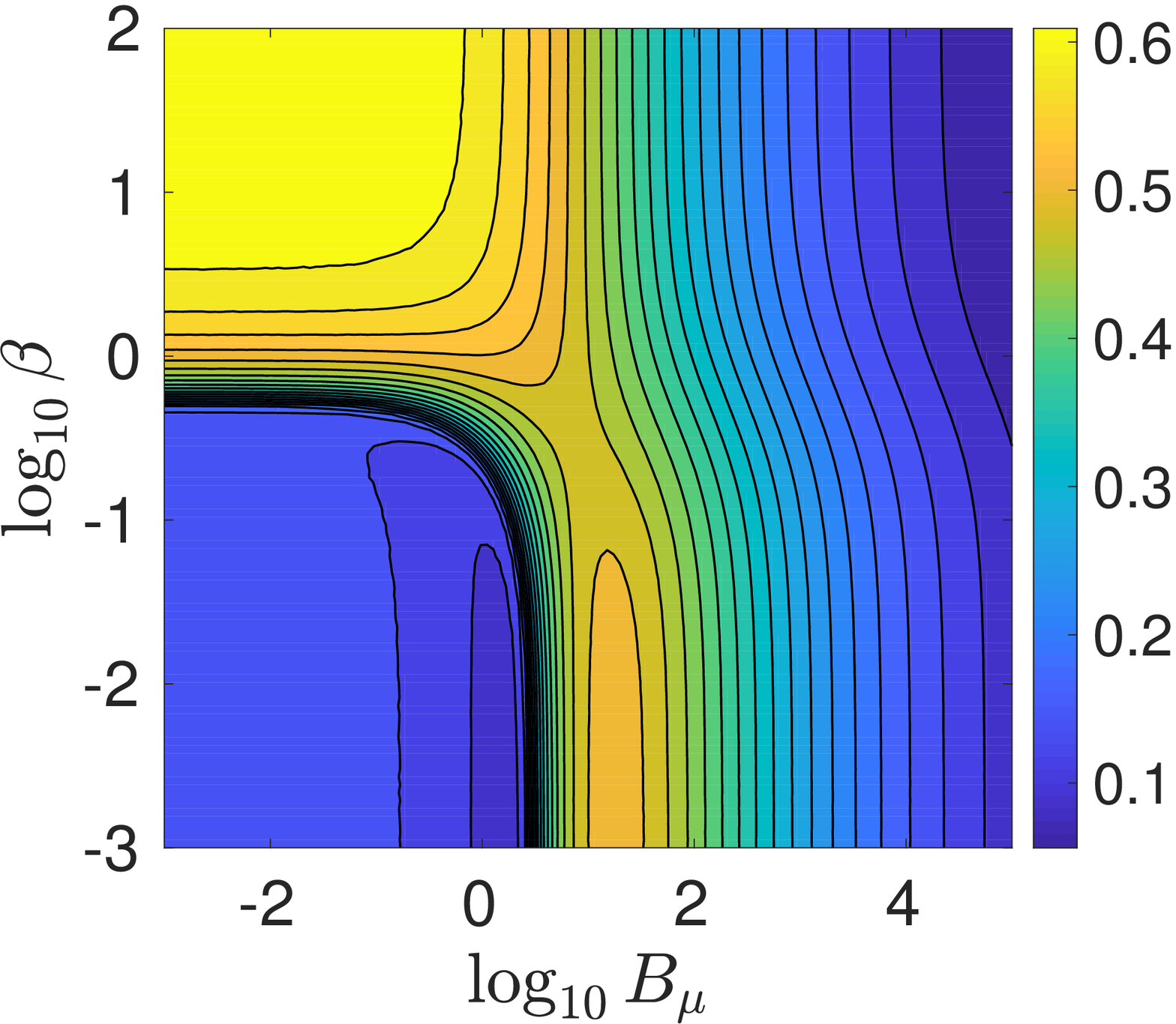}}
 \end{picture}
 \begin{picture}(0,0)
 \put(30,147.5){\includegraphics[height=2.8cm]{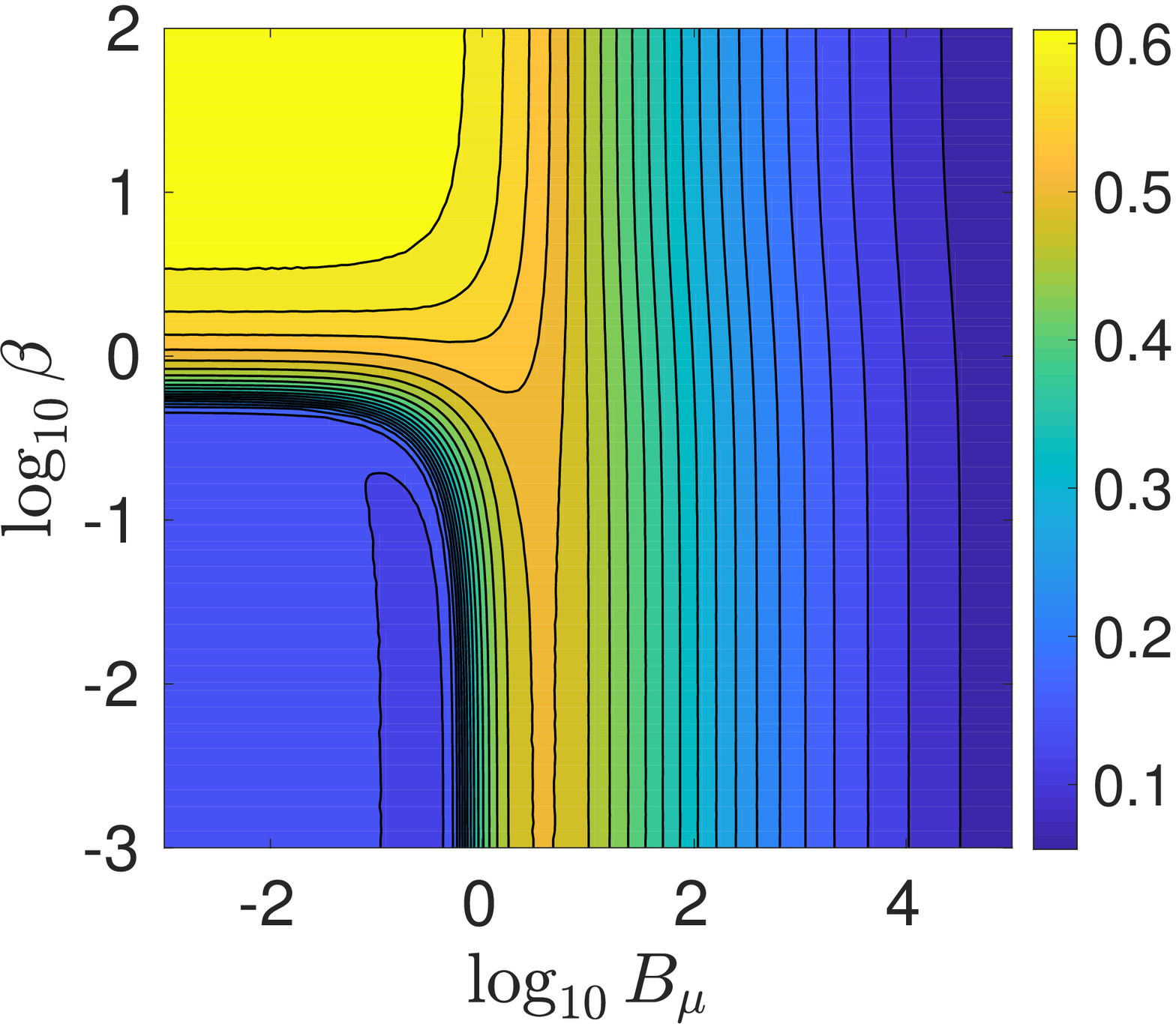}}
 \end{picture}
 \caption{(Colour online) ($a$--$c$) Maximum growth rate, $\omega_m$, and ($d$--$f$) corresponding wavenumber, $k_m$, as functions of $\Bou_{\mu}$ for $\Reycap=0.01$ and several values of $\beta$ indicated in the legend, for (a,d) $\mu_{s0}/\kappa_{s0}=0.1$ ($b$,$e$) $\mu_{s0}/\kappa_{s0}=1$ and ($c$,$f$) $\mu_{s0}/\kappa_{s0}=10$. The insets show the contours of constant $\omega_m$ ($a$--$c$) and of constant $k_m$ ($d$--$f$) in the $(\Bou_{\mu},\beta)$ parameter plane.\label{fig:figure5}}
 \end{figure}

\begin{figure}
 \includegraphics[width=\textwidth]{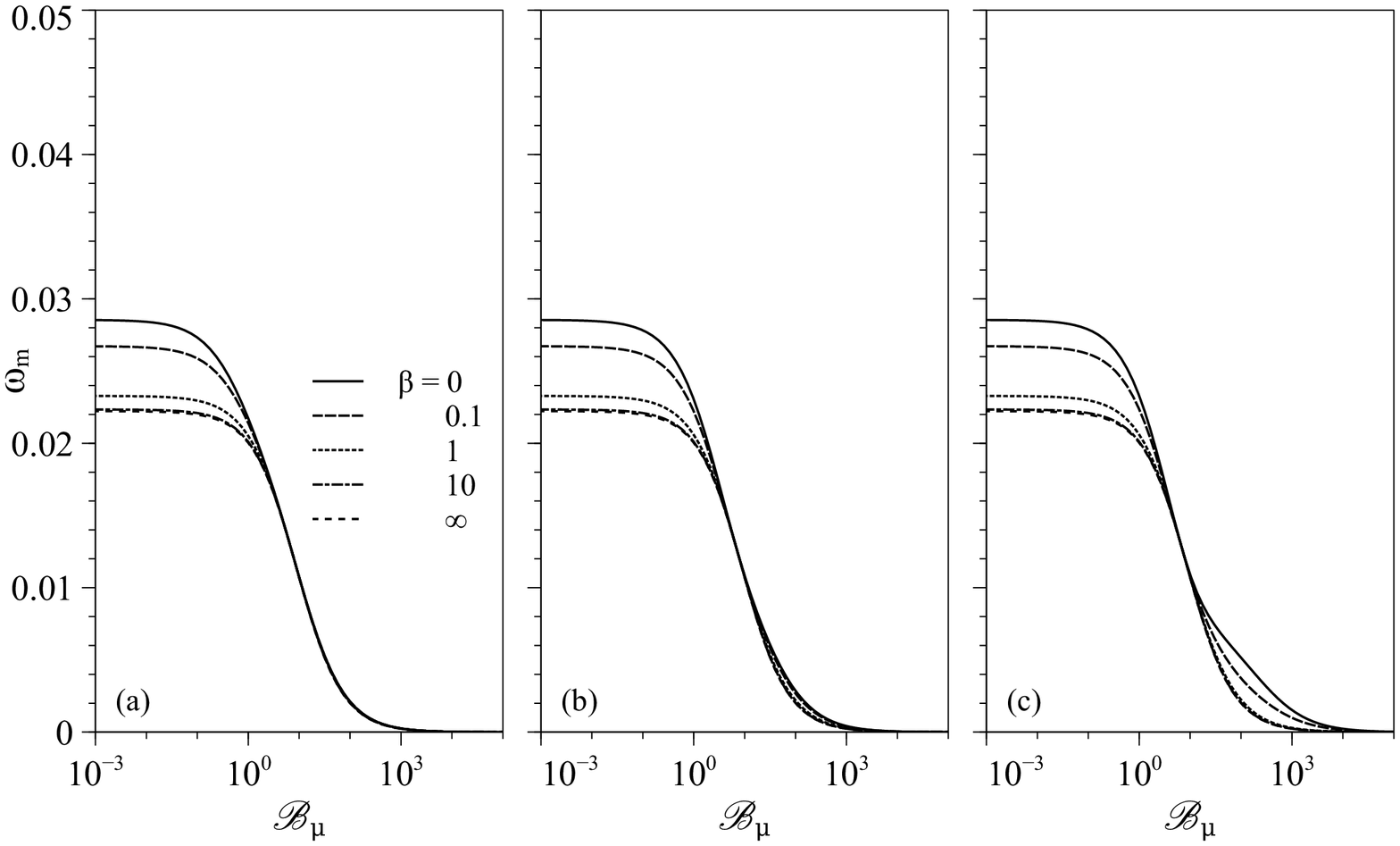}%
 \begin{picture}(0,0)
 \put(-115,137.5){\includegraphics[height=2.8cm]{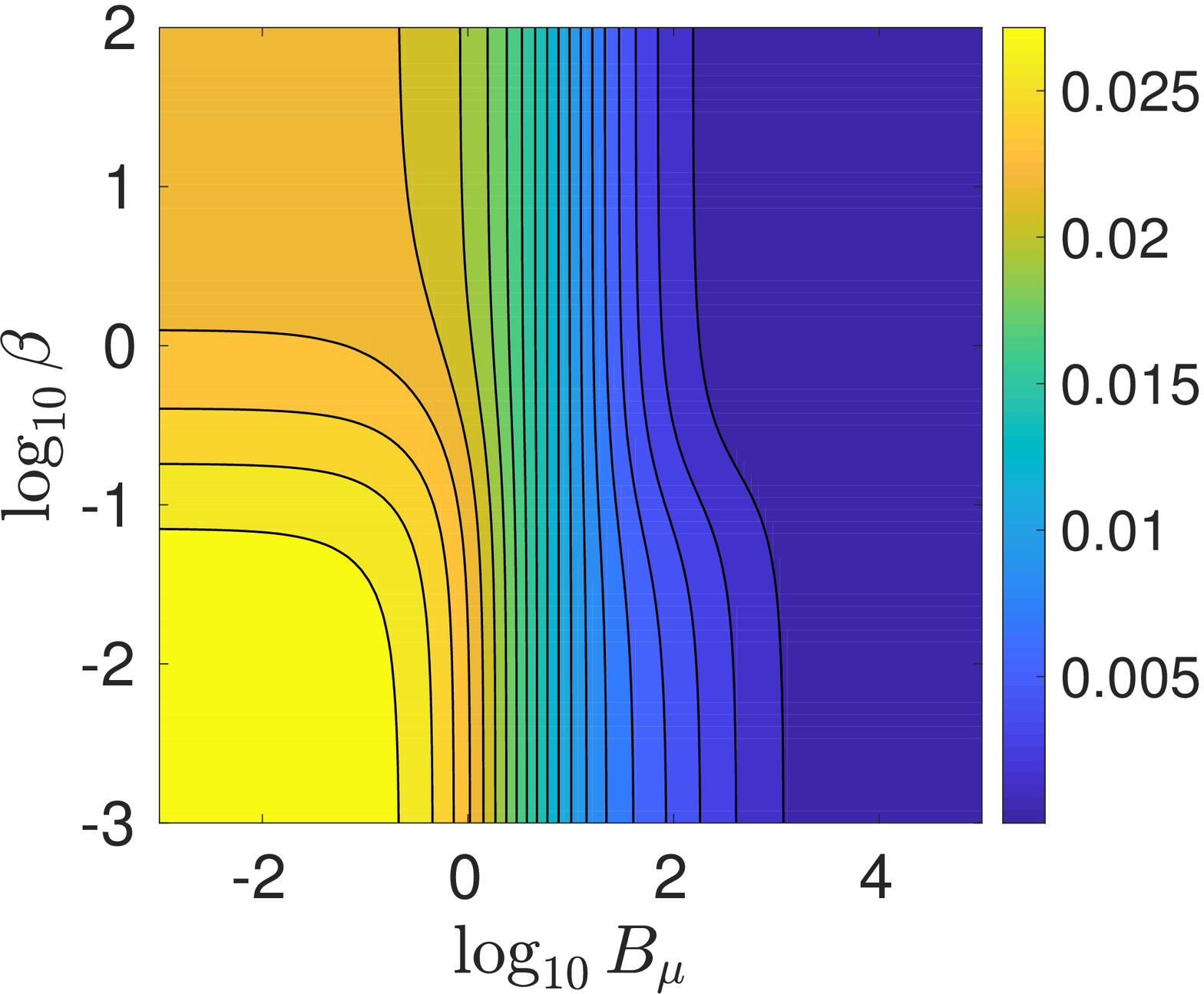}}
 \end{picture}
 \begin{picture}(0,0)
 \put(150,147.5){\includegraphics[height=2.8cm]{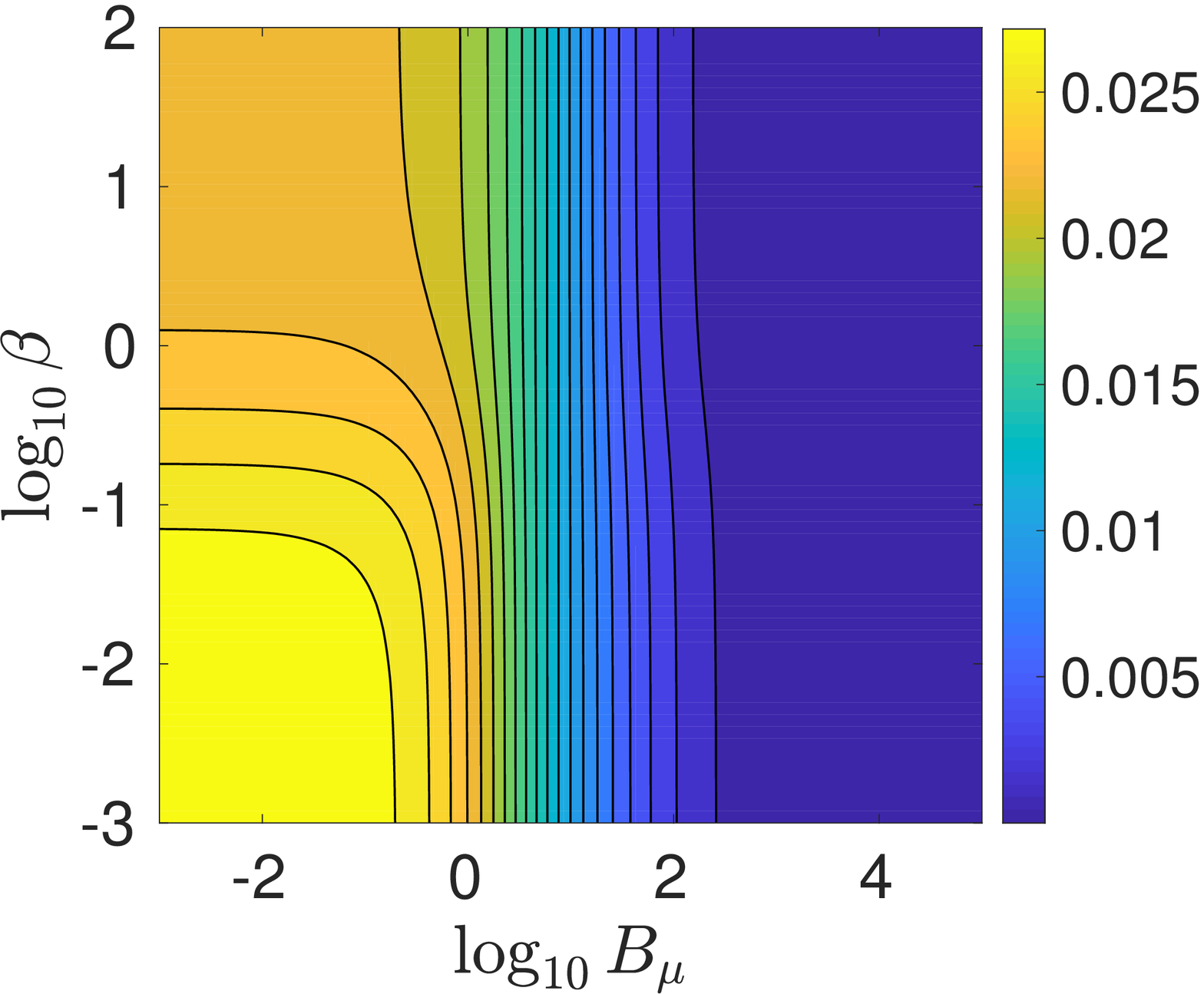}}
 \end{picture}
 \begin{picture}(0,0)
 \put(30,147.5){\includegraphics[height=2.8cm]{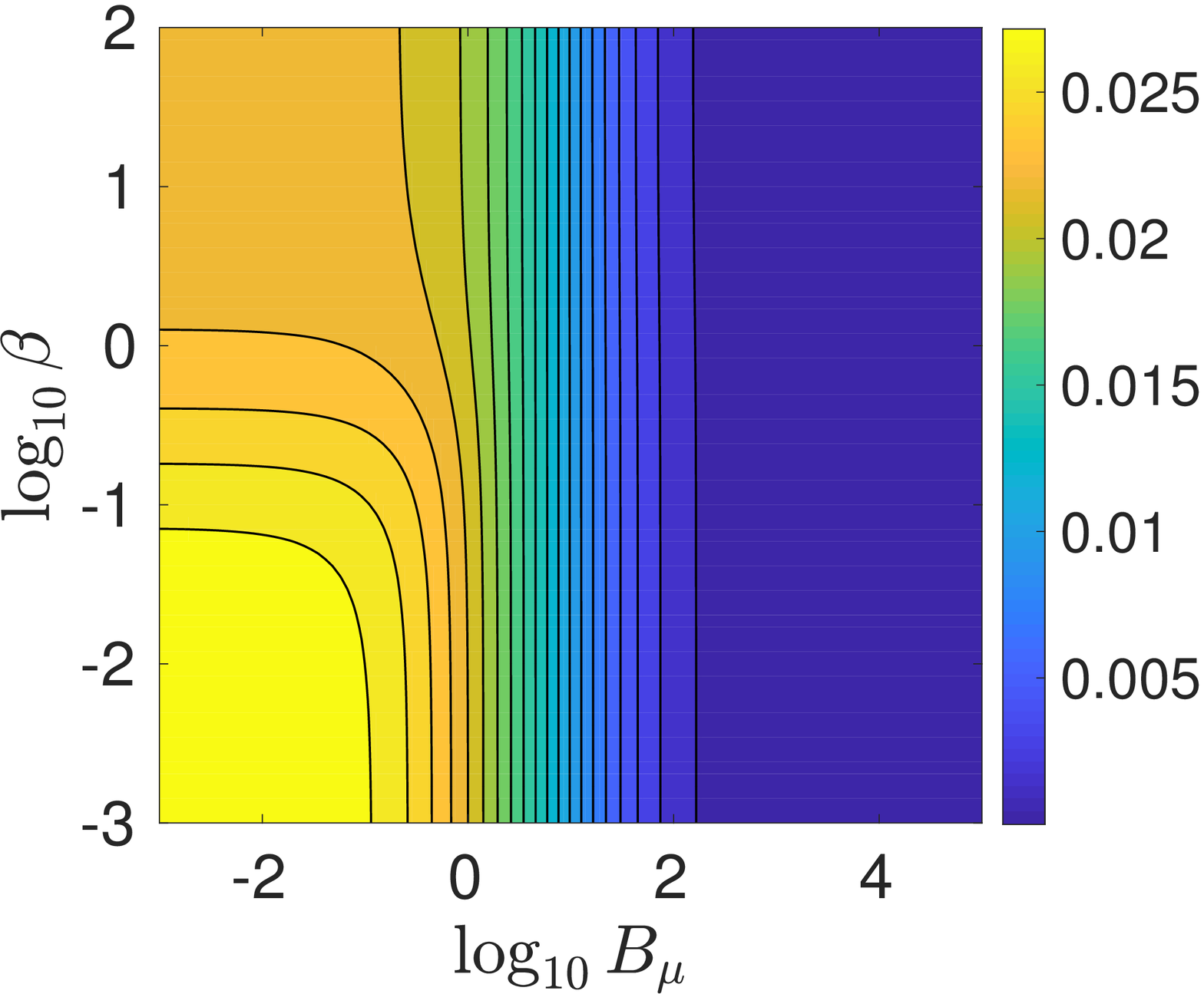}}
 \end{picture}\\
 \includegraphics[width=\textwidth]{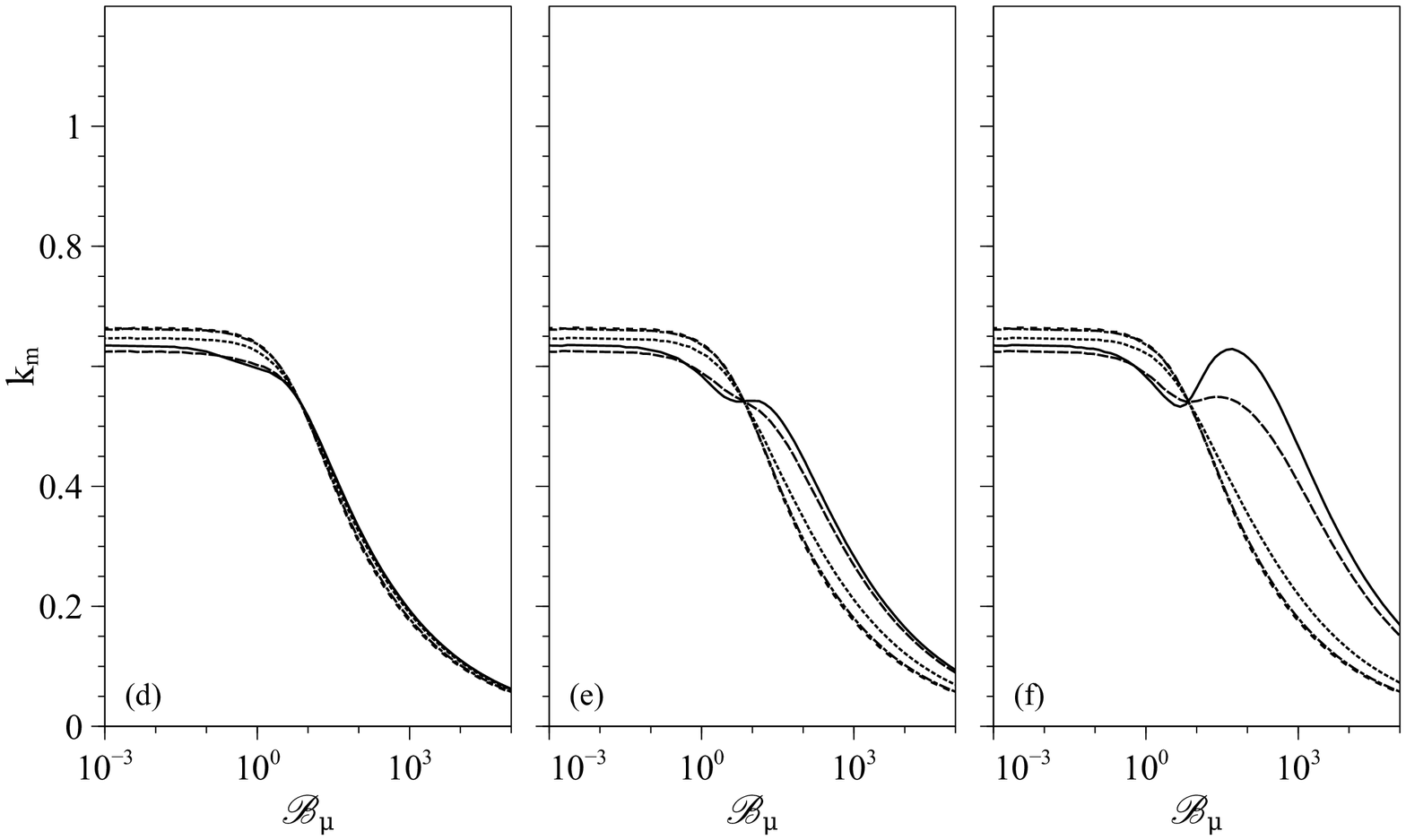}%
 \begin{picture}(0,0)
 \put(-115,137.5){\includegraphics[height=2.8cm]{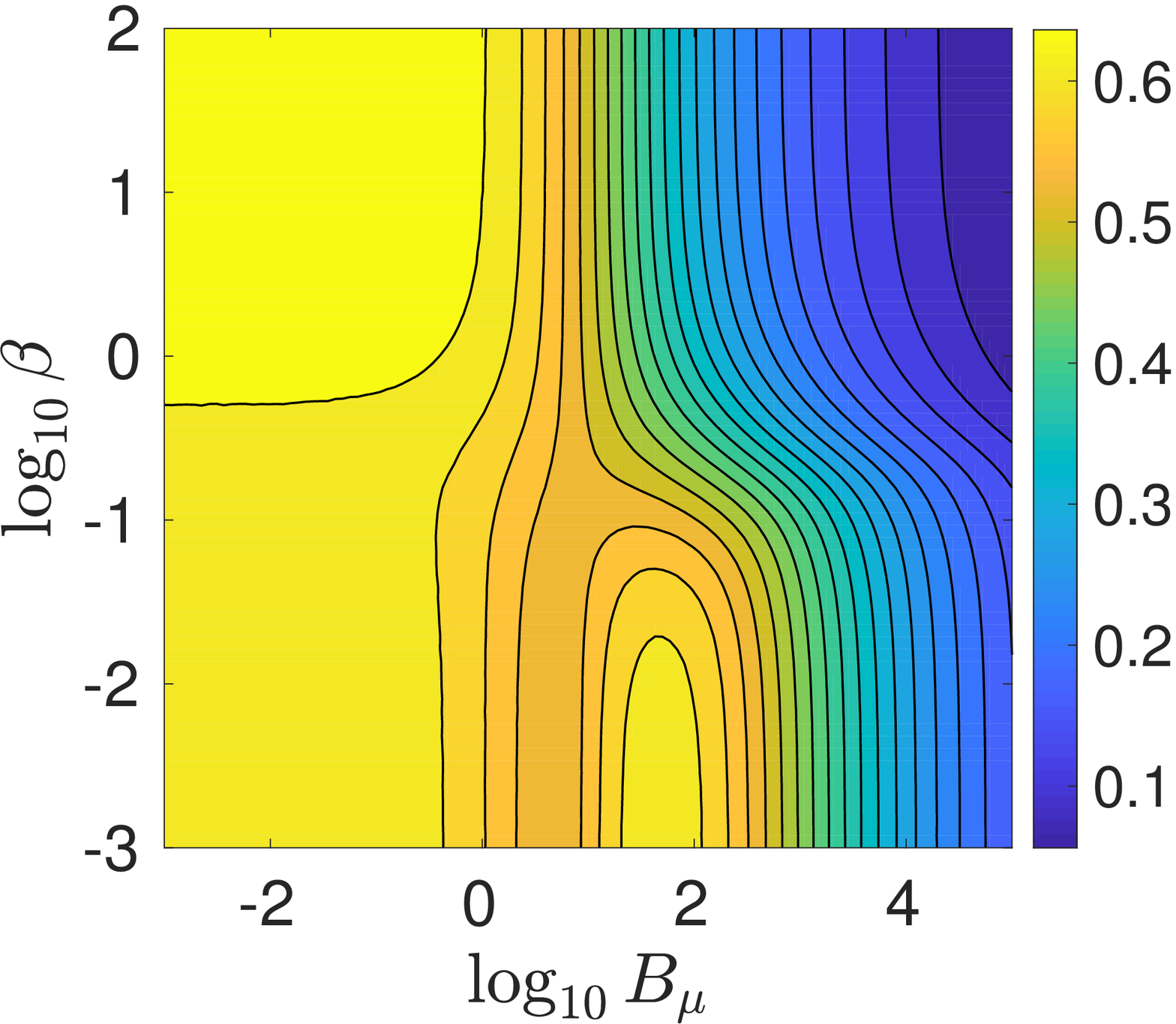}}
 \end{picture}
 \begin{picture}(0,0)
 \put(150,147.5){\includegraphics[height=2.8cm]{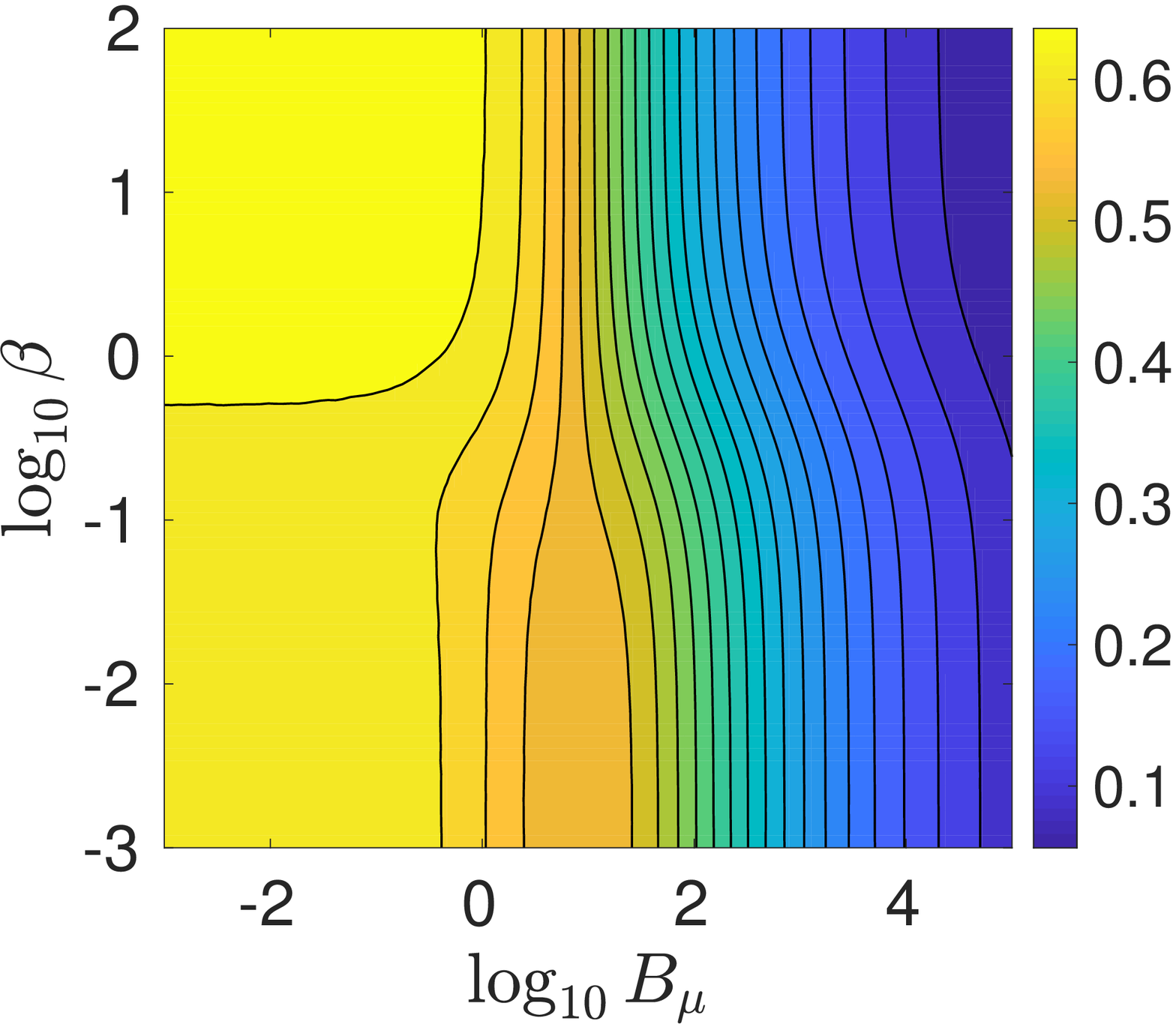}}
 \end{picture}
 \begin{picture}(0,0)
 \put(30,147.5){\includegraphics[height=2.8cm]{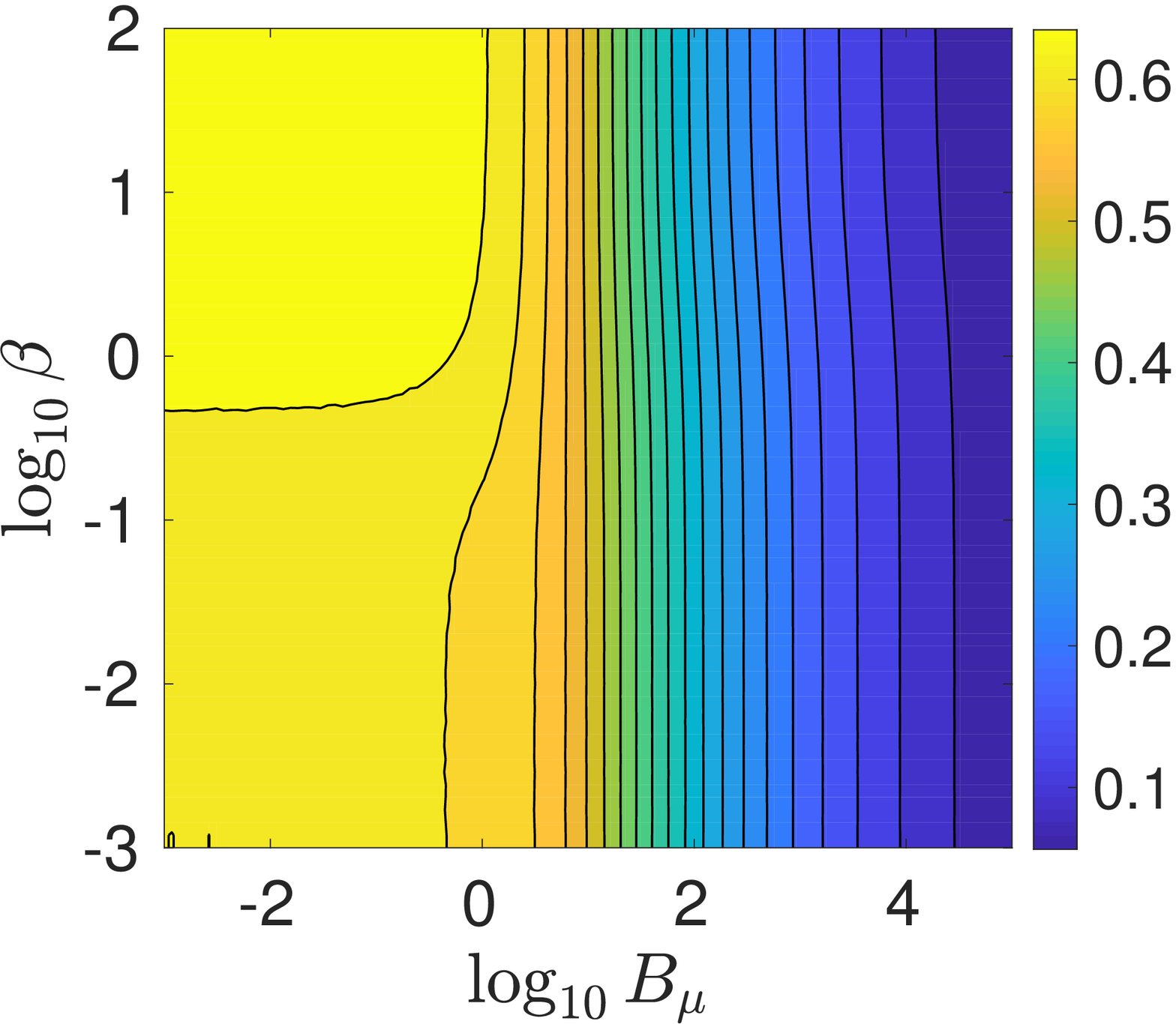}}
 \end{picture}
 \caption{(Colour online) Same as figure~\ref{fig:figure5} for $\Reycap=100$.\label{fig:figure6}}
 \end{figure}

To illustrate the parametric dependence of $\omega_m$ and $k_m$ we have computed figures~\ref{fig:figure5} and~\ref{fig:figure6}, where the maximum growth rate, $\omega_m$ ($a$--$c$), and the corresponding wavenumber, $k_m$ ($d$--$f$), are plotted as functions of $\Bou_{\mu}$ for several values of $\beta$ indicated in the legends, and three different values of the surface viscosity ratio, namely $\mu_{s0}/\kappa_{s0}=0.1$, 1 and 10 in panels ($a$,$d$), ($b$,$e$) and ($c$,$f$), respectively. The value of $\Reycap=0.01$ in figure~\ref{fig:figure5}, while $\Reycap=100$ in figure~\ref{fig:figure6}. The insets show isocontours of $\omega_m$ ($a$--$c$) and of $k_m$ ($d$--$f$), in the $(\Bou_{\mu},\beta)$ parameter plane. Firstly, as already mentioned, when $\Bou_{\mu} \to \infty$, $\omega_m \to 0$ due to the dissipative nature of the surface viscosities, in contrast with the finite value of $\omega_m$ reached when $\beta \to \infty$~\citep{Timmermans02}. It can also be observed in figures~\ref{fig:figure5}($a$-$c$) and~\ref{fig:figure6}($a$-$c$) that, for $\beta \lesssim 1$, $\omega_m$ decreases faster when $\mu_{s0}/\kappa_{s0} \ll 1$. Hence, when both surface viscosities increase and $\kappa_{s0} > \mu_{s0}$, the growth rate of perturbations is smaller, as already deduced from figures~\ref{fig:figure2}($a$,$b$) and~\ref{fig:figure3}($a$,$b$), and also by~\citet{Whitaker76} in the limit $\Reycap \gg 1$.

With regard to $k_m$, and starting with the influence of $\beta$ in the limit $\Bou_{\mu}\to 0$, our results reveal that $k_m$ first slightly decreases for $0 \leq \beta \lesssim 0.1$, and that an increase in $\beta$ upon this limit, i.e. $\beta\gtrsim O(1)$, implies a sudden increase in $k_m$. As shown by figures~\ref{fig:figure5}($d$-$f$) and~\ref{fig:figure6}($d$--$f$), the former effect is more pronounced for $\Reycap \gg 1$, and the latter effect, which was already highlighted by~\citet{Timmermans02}, is more noticeable for $\Reycap \ll 1$.

Concerning the effect of surface viscosities on $k_m$, figure~\ref{fig:figure5}($d$--$f$) for $\Reycap = 0.01$ shows that, when $\beta \lesssim 0.3$ and $\Bou_{\mu}$ increases for a fixed value of $\mu_{s0}/\kappa_{s0}$, there is a slight decrease of $k_m$. As $\mu_{s0}/\kappa_{s0}$ increases, the latter phenomenon is more pronounced and occurs at higher values of $\Bou_{\mu}$. 
When $\Bou_{\mu}$ increases past these limits, there is a sudden growth of $k_m$, which is more noticeable for higher values of $\mu_{s0}/\kappa_{s0}$, also taking place in a larger range of $\Bou_{\mu}$. 
Hence, for these intervals of $\Bou_{\mu}$ surface viscosity provides an alternative mechanism for short-wavelength selection, which can also be seen in figure~\ref{fig:figure2}($a$,$b$). Furthermore, when the maximum value of $k_m$ is reached, an increase of $\Bou_{\mu}$ decreases $k_m$ monotonically. On the contrary, when $\Reycap \gg 1$, this non-monotonic behaviour only appears when $\beta \ll 1$ and $\mu_{s0}/\kappa_{s0} \gg 1$, i.e. when the flow is dominated by the shear viscosity, as depicted by figures~\ref{fig:figure6}($d$--$f$). Indeed, for values of $\beta \gtrsim 1$, the dependence of $k_m$ on $\Bou_{\mu}$ is monotonically decreasing, independently of $\Reycap$ and $\mu_{s0}/\kappa_{s0}$, as shown in figures~\ref{fig:figure5}($d$--$f$) and~\ref{fig:figure6}($d$--$f$).


Another interesting feature observed in figures~\ref{fig:figure5}($d$--$f$) and~\ref{fig:figure6}($d$--$f$) is that there is a certain value of $\Bou_{\mu}$ at which $\omega_m$ and the corresponding $k_m$ are independent of $\beta$, since all the curves intersect at this point. For $\Reycap = 0.01$ this value is approximately $\Bou_{\mu} \sim 8.1$, and for $\Reycap = 100$, $\Bou_{\mu} \sim 6.8$. Above these values, the dependence of $k_m$ on $\beta$ changes, since an increase of $\beta$ reduces the value of $k_m$. Additionally, past these values of $\Bou_{\mu}$ and for $\mu_{s0}/\kappa_{s0} \lesssim 1$, every $\omega_m$ curve collapses as $\Bou_{\mu}$ increases, indicating that $\omega_m$ becomes independent of $\beta$. On the contrary, for $\mu_{s0}/\kappa_{s0} \gtrsim 1$ and $\beta \lesssim 1$, the maximum growth rate, $\omega_m$, is higher than in the latter case for the same value of $\Bou_{\mu}$, which can also be noticed in figures~\ref{fig:figure2}($a$,$b$) and~\ref{fig:figure3}($a$,$b$) for $\Bou_{\mu} = 50$, with a more pronounced effect for $\Reycap \gg 1$.

Finally, the values of the surface shear and dilatational Boussinesq numbers may seem unrealistically overestimated in figures~\ref{fig:figure5} and~\ref{fig:figure6}. However, if we take as an example a liquid thread of 400~$\mu$m radius, and an aqueous octanoic acid solution with a bulk concentration of $9 \times 10^{-5}$ mol \, ml$^{-1}$ and the bulk properties of water at 25$^{\circ}$C, the surface shear viscosity takes the value of $\mu_s = 4 \times 10^{-5}$ Pa$\,$s$\,$m, and the dilatational viscosity $\kappa_s = 2.7 \times 10^{-4}$ Pa$\,$s$\,$m~\citep{Ting1984}, providing values of the Boussinesq numbers of $\Bou_{\mu} = 100$ and $\Bou_{\kappa} = 675$, and thus a ratio $\mu_{s}/\kappa_{s}=0.148$. Since the liquid is water, $\Reycap \gg 1$ and thus this example would correspond approximately to figures~\ref{fig:figure6}(a) and (d), with a maximum growth rate of $\omega_m \sim 2 \times 10^{-3}$ and its corresponding wavenumber $k_m \sim 0.33$. Indeed, the surface shear Boussinesq number for this solution can be higher; for example, if we consider a concentration of $4.5 \times 10^{-5}$ mol \, ml$^{-1}$, $\Bou_{\mu} = 150$, $\Bou_{\kappa} = 950$, and thus $\mu_s/\kappa_s = 0.158$, $\omega_m \sim 1.46 \times 10^{-3}$ and $k_m \sim 0.3$. Therefore, these two realistic examples show that the values of $\Bou_{\mu}$ and $\Bou_{\kappa}$ may be large enough to produce a substantial decrease in $\omega_m$.

\section{Long-wave expansion \label{sec:longwave}}

Despite the usefulness of the linear stability analysis presented in~\S\ref{sec:results}, it cannot describe the nonlinear dynamics of the liquid column, namely the final stages of its breakup and the possible formation of satellite droplets. Besides, even in the axisymmetric case, integrating the complete Navier--Stokes equations in the presence of surfactant molecules is computationally expensive. Hence, the present section is devoted to derive two different 1D approximations that take into account the effects of Marangoni stresses and surface viscosities when the liquid column is covered with insoluble surfactant. It is well known that many features of the dynamics of viscous liquid jets and bridges can be accurately described using this long-wave limit, both in the linear regime and close to breakup with and without the presence of surfactant~\citep[see e.g.][]{Eggers1993,Craster2002, Timmermans02,Craster2009}. Here, following the same spirit as~\citet{GyC}, we consider the dimensionless parameter $\varepsilon = R/L$, where $L$ is the characteristic axial length of the liquid thread. Assuming that $L \gg R$, $\varepsilon \ll 1$ is the appropriate small parameter to derive the 1D models. To this end, it proves convenient to make the flow variables dimensionless using the following characteristic scales:
\begin{subequations}
\begin{align}
& r_c = z_c = \frac{R}{\varepsilon}, \, \, t_c = \frac{\mu R}{\varepsilon \, \sigma_0}, \, \, u_c = w_c = \frac{\sigma_0}{\mu}, \, \, a_c = R, & \\
& p_c = \frac{\sigma_0}{R}, \, \, \sigma_c = \sigma_0, \, \, \Gamma_c = \Gamma_0, \, \, \mu_{s,c} = \mu_{s0}, \, \, \kappa_{s,c} = \kappa_{s0}
\end{align}
\end{subequations}
Hereafter, for simplicity, the dimensionless variables have the same notation as the dimensional ones used in~\S\ref{sec:formulation} and~\S\ref{sec:results}.

The regularity of the solution at $r = 0$ together with the axisymmetry condition require that the pressure and the axial velocity fields have to be even functions of $r$, whilst the radial velocity field has to be an odd function of $r$, suggesting the following expansion in the radial coordinate $r \sim \varepsilon$~\citep{EggersDupont,GyC},
\begin{equation}\label{eq:w_expansion}
w(r,z,t) = w_0(z,t) + \frac{1}{2} r^2  w_2(z,t) + ... + \frac{1}{(2j)!}r^{2j}  w_{2j},
\end{equation}
\begin{equation}\label{eq:p_expansion}
p(r,z,t) = p_0(z,t) + \frac{1}{2} r^2  p_2(z,t) + ... + \frac{1}{(2j)!}r^{2j}  p_{2j},
\end{equation}
\begin{equation}\label{eq:u_expansion}
u(r,z,t) = - \frac{1}{2} r w_0'(z,t) - \frac{1}{8} r^3 w_2'(z,t) - ... - \frac{2j+1}{(2j+2)!} r^{2j +1} w'_{2j},
\end{equation}
for $j \in \mathbb{N}$ and~\eqref{eq:u_expansion} satisfying the continuity equation~\eqref{eq:continuity}.

\subsection{Leading-order model}\label{subsec:leadingorder}

Introducing~\eqref{eq:w_expansion}--\eqref{eq:u_expansion} into the dimensionless versions of~\eqref{eq:kinematic} and~\eqref{eq:zmomentum}, the leading-order kinematic condition and the axial momentum equation read, respectively,
\begin{equation}
\frac{\partial a^2}{\partial t} + \frac{\partial(a^2 w_0)}{\partial z} + O(\varepsilon^2) = 0,
\end{equation}
\begin{equation}\label{eq:leadordermom}
\Reycap \left(\frac{\partial w_0}{\partial t} +  w_0 w_0' \right) = -p_0' +  \varepsilon (w_0'' + 2 w_2) + O(\varepsilon^2),
\end{equation}
where $p_0$ and $w_2$ can be obtained from~\eqref{eq:bc_normal} and~\eqref{eq:bc_tangential}, respectively, which at leading order yield the following normal and tangential stress balances at $r = \varepsilon a$:
\begin{equation}
p_0 = - \varepsilon w_0' + \mathcal{C} \left[\sigma + \varepsilon \frac{(\Bou_{\kappa} \kappa_s - \Bou_{\mu} \mu_s)}{a} \left( (a w_0)' - \frac{\mathcal{C} (a^2 w_0)'}{2}  \right) \right] - \varepsilon \frac{\Bou_{\mu} \mu_s w_0'}{a} + O(\varepsilon^2),
\end{equation}
\begin{align}
& w_2 = \frac{\sigma'}{\varepsilon a} + \frac{w_0''}{2} + \frac{3 a' w_0'}{a} + \frac{1}{a} \frac{\partial}{\partial z} \left[\frac{(\Bou_{\kappa} \kappa_s - \Bou_{\mu} \mu_s)}{a}\left((aw_0)' -  \frac{\mathcal{C}  (a^2 w_0)'}{2} \right) \right] \nonumber & \\
& + \frac{2 \Bou_{\mu} (\mu_s a w_0')'}{a^2} + \frac{\Bou_{\mu} \mu_s a' w_0'}{a^2} + O(\varepsilon^2).
\end{align}
Additionally, the transport equation~\eqref{eq:surftransport} for the surfactant concentration, $\Gamma(z,t)$, reads, at leading order, 
\begin{equation}\label{eq:1Dsurfactant}
\frac{\partial \Gamma}{\partial t} + \frac{1}{a}\frac{\partial (a w_0 \Gamma )}{\partial z} - \frac{\mathcal{C} \Gamma}{2a}\frac{\partial(a^2 w_0)}{\partial z} + O(\varepsilon^2) = 0.
\end{equation}
Although it does not affect the linear regime,~\eqref{eq:1Dsurfactant} retains the full curvature, $\mathcal{C}$, unlike the leading-order expressions for the surfactant concentration derived by~\citet{Timmermans02},\citet{Liao2006} and~\citet{Craster2009}, which include the approximation $\mathcal{C} \simeq a^{-1}$.
Finally, 
eliminating the small parameter $\varepsilon$ from the formulation via the substitutions $z \rightarrow \varepsilon z$, $t \rightarrow \varepsilon t$, and $w_{2j} \rightarrow w_{2j}/\varepsilon^{2j}$, the leading-order 1D model consists of the following three coupled equations:
\begin{equation}\label{eq:1Dcontinuity}
\frac{\partial a^2}{\partial t} + \frac{\partial(a^2 w_0)}{\partial z} = 0,
\end{equation}
\begin{align}\label{eq:1Dmomentum_dimless}
&  \Reycap \left(\frac{\partial w_0}{\partial t} + w_0 \frac{\partial w_0}{\partial z} \right) =  \frac{3}{a^2} \frac{\partial}{\partial z} \left( a^2 \frac{\partial  w_0}{\partial z} \right)  \nonumber & \\
& - \frac{\partial}{\partial z}\left\{\mathcal{C} \left[\sigma + \frac{(\Bou_{\kappa} \kappa_s -\Bou_{\mu} \mu_s )}{a}\left( \frac{\partial (a  w_0)}{\partial z} - \frac{\mathcal{C}}{2}\frac{\partial(a^2  w_0)}{\partial z} \right) \right] \right\} \nonumber & \\
& + \frac{2}{a} \frac{\partial}{\partial z}\left[\sigma + \frac{(\Bou_{\kappa} \kappa_s - \Bou_{\mu} \mu_s)}{a} \left( \frac{\partial (a w_0) }{\partial z} - \frac{\mathcal{C}}{2} \frac{\partial (a^2 w_0)}{\partial z} \right) \right] + \frac{5 \Bou_{\mu} }{a^2} \frac{\partial }{\partial z}\left(\mu_s a \frac{\partial w_0}{\partial z} \right),
\end{align}
\begin{equation}\label{eq:1Dsurfactant2}
\frac{\partial \Gamma}{\partial t} + \frac{1}{a}\frac{\partial (a w_0 \Gamma)}{\partial z} - \frac{\mathcal{C} \Gamma}{2a}\frac{\partial(a^2 w_0)}{\partial z} = 0.
\end{equation}
These must be complemented with three equations of state, $\sigma(\Gamma)$, $\mu_s(\Gamma)$ and $\kappa_s(\Gamma)$, relating the surface tension coefficient and the two surface viscosities with the surfactant concentration. However, note that explicit expressions for the three equations of state are not needed in the linearised analysis presented herein.
Equations~\eqref{eq:1Dcontinuity}--\eqref{eq:1Dsurfactant2} recover the limit when only gradients of $\sigma$ are considered by setting $\Bou_{\mu} \to 0$ and $\Bou_{\kappa} \to~0$~\citep{Timmermans02, Liao2006, Craster2009}, and also the limit of a clean liquid thread when $\Gamma \to 0$, $\sigma \to 1$, $\Bou_{\mu} \to 0$ and $\Bou_{\kappa} \to 0$~\citep{EggersDupont,GyC}.

\subsection{Parabolic model}\label{subsec:parabolic}

The accuracy of the 1D approximation deduced in~\S\ref{subsec:leadingorder} can be improved by retaining terms of order $O(\varepsilon^2)$, leading to the so-called \textit{parabolic model}, whose relative error is of order $O(\varepsilon^4)$~\citep{GyC}. At order $O(\varepsilon^2)$, the kinematic condition, the surfactant transport equation and the axial momentum equation read, respectively,
\begin{equation}
\frac{\partial a^2}{\partial t} + \frac{\partial(a^2 w_0)}{\partial z} + \frac{\varepsilon^2}{4} \frac{\partial (a^4 w_2)}{\partial z} + O(\varepsilon^4) = 0.
\end{equation}
\begin{align}
& \frac{\partial \Gamma}{\partial t} + \frac{(a  w_0 \Gamma)'}{a} - \frac{ \mathcal{C} (a^2 w_0)' \Gamma}{2 a}+ \frac{\varepsilon^2}{2 a} \biggl[ (a^3 w_2 \Gamma)' - a'^2 (a w_0)' \Gamma - (a a' (a w_0)')' \Gamma  \nonumber & \\
& - \frac{\mathcal{C} \, \Gamma}{2} \left( \frac{(a^4 w_2)'}{2} - a'^2 (a^2 w_0)' \right) \biggr] + O(\varepsilon^4) = 0,
\end{align}
\begin{equation}\label{eq:second_order_zmomentum}
\Reycap \left( \frac{\partial w_2}{\partial t} + w_0 w_2' \right) = -p_2' + \varepsilon \left(w_2'' + \frac{4 w_4}{3} \right) + O(\varepsilon^2).
\end{equation}
Here $w_4(z,t)$ can be obtained from the second-order truncation of the tangential stress balance,
\begin{align}
& a \left(w_2 - \frac{w_0''}{2} \right) - 3 a' w_0' + \varepsilon^2 \left(\frac{a^3 w_4}{6} -\frac{a^3 w_2''}{8} - \frac{7 a^2 a' w_2'}{4} - \frac{3 a a'^2 w_2}{2} + \frac{3 a a'^2 w_0''}{4} \right. \nonumber & \\
& \left.  + \frac{3 a'^3 w_0'}{2}\right) = \frac{\sigma'}{\varepsilon} + \frac{\partial }{\partial z} \left[\frac{(\Bou_{\kappa} \kappa_s - \Bou_{\mu} \mu_s )}{a} \left((a w_0)' -  \frac{\mathcal{C} (a^2 w_0)'}{2} \right) \right] + \frac{2 \Bou_{\mu} (\mu_s a w_0')'}{a}  \nonumber & \\
&  + \frac{\Bou_{\mu} \mu_s a' w_0'}{a} + \varepsilon^2 \frac{\partial}{\partial z}\left[\frac{(\Bou_{\kappa} \kappa_s- \Bou_{\mu} \mu_s)}{2 a} \biggl( (a^3 w_2)' - a'^2 (a w_0)' - (a a' (a w_0)')' \right.   \nonumber & \\
&  \left.    - \frac{\mathcal{C}}{2} \left( \frac{(a^4 w_2)'}{2} - a'^2 (a^2 w_0)' \right) \biggr) \right] + \varepsilon^2 \frac{\Bou_{\mu} a'^2 ( \mu_s a w_0')'}{a} + \frac{\varepsilon^2 \Bou_{\mu} }{a} \frac{\partial}{\partial z} \biggl[ \mu_s a ( (a^2 w_2)'   \nonumber & \\
& - 4 a'^2 w_0' - a a' w_0'') \biggr] +  \varepsilon^2 \Bou_{\mu} \mu_s a' \left( \frac{a w_2'}{4} + 2 a'' w_0' \right) + O(\varepsilon^4),
\end{align}
and $p_2(z,t)$ can be deduced from the first-degree approximation in $\varepsilon$ of the radial momentum equation~\eqref{eq:rmomentum},
\begin{equation}
p_2 = \Reycap \left( \frac{1}{2}\frac{\partial w_0'}{\partial t} + \frac{w_0 w_0''}{2} - \frac{w_0'^2}{4} \right) - \frac{\varepsilon}{2}(w_0''' + 2 w_2') + O(\varepsilon^3).
\end{equation}
Equation~\eqref{eq:second_order_zmomentum} is coupled to~\eqref{eq:leadordermom}, i.e. the first-order momentum equation for $w_0(z,t)$, whose unique unknown is $p_0(z,t)$. A higher-order approximation of $p_0$ can be obtained from the second-order normal stress balance,
\begin{align}
& p_0 + \varepsilon^2 \frac{a^2 p_2}{2} = \mathcal{C} \sigma + \varepsilon \left[-w_0' + \frac{\mathcal{C} (\Bou_{\kappa} \kappa_s  - \Bou_{\mu} \mu_s )}{a} \left((a w_0)' -  \frac{\mathcal{C} (a^2 w_0)'}{2} \right)  - \frac{\Bou_{\mu} \mu_s w_0'}{a} \right.  \nonumber & \\
& \left.  + \varepsilon^2\left(3 a'^2 w_0' - \frac{3 a^2 w_2'}{4} - 2 a a'w_2 + a a'w_0''\right) + \varepsilon^2 \frac{\mathcal{C} (\Bou_{\kappa} \kappa_s - \Bou_{\mu} \mu_s ) }{2 a} \biggl( (a^3 w_2)'  - a'^2 (a w_0)'  \right.   \nonumber & \\
& \left.  - (a a'(a w_0)')'   - \frac{\mathcal{C}}{2} \left( \frac{(a^4 w_2)'}{2} - a'^2 (a^2 w_0)' \right) \biggr)  -  \varepsilon^2 \Bou_{\mu} \mu_s \left( \frac{a w_2'}{4} + 2 a'' w_0' - \frac{a'^2 w_0'}{2 a} \right) \right] + O(\varepsilon^4).
\end{align}
Hence, introducing the expressions for $p_2$ and $w_4$ into~\eqref{eq:second_order_zmomentum}, and those for $p_0$ and $p_2$ into~\eqref{eq:leadordermom}, and eliminating the small parameter $\varepsilon$ from the formulation via the substitutions $z \rightarrow \varepsilon z$, $t \rightarrow \varepsilon t$ and $w_{2j} \rightarrow w_{2j}/\varepsilon^{2j}$, the four equations of the parabolic model read
\begin{equation}\label{eq:secondorder_continuity}
\frac{\partial a^2}{\partial t} + \frac{\partial(a^2 w_0)}{\partial z} + \frac{1}{4} \frac{\partial (a^4 w_2)}{\partial z} = 0,
\end{equation}
\begin{align}\label{eq:secondorder_w0}
&  \Reycap \left( \frac{\partial w_0}{\partial t} + w_0 w_0' - \left[ \frac{a^2}{4} \left(\frac{\partial w_0'}{\partial t} + w_0 w_0'' - \frac{w_0'^2}{2} \right) \right]' \right) =  2w_0'' + 2 w_2 -6 a'a''w_0'  \nonumber & \\
&  -(4 a'^2 + a a'')w_0'' - \frac{3 a a' w_0'''}{2} - \frac{a^2 w_0''''}{4} + 2(a'^2 + a a'')w_2  + \frac{5 a a' w_2'}{2} + \frac{a^2 w_2''}{4}  \nonumber & \\
&  - \frac{\partial}{\partial z} \left[ \mathcal{C} \sigma + \frac{\mathcal{C} (\Bou_{\kappa} \kappa_s - \Bou_{\mu} \mu_s)}{a} \left((a w_0)'  + \frac{(a^3 w_2)' - a'^2(a w_0)' - (a a' (a w_0)' )'}{2}   \right. \right.  \nonumber & \\
& \left. \left.  - \frac{\mathcal{C}}{2} \left( (a^2 w_0)' + \frac{(a^4 w_2)'}{4} - \frac{a'^2 (a^2 w_0)' }{2} \right) \right) - \Bou_{\mu} \mu_s \left( \frac{w_0'}{a} + \frac{a w_2'}{4} + 2 a'' w_0' - \frac{a'^2 w_0'}{2 a}  \right) \right],
\end{align}
\begin{align}\label{eq:secondorder_w2}
&  \Reycap \left( \frac{\partial w_2}{\partial t} + w_0 w_2' + \frac{w_0 w_0'''}{2} + \frac{1}{2}\frac{\partial w_0''}{\partial t} \right)  = \frac{4 w_0''}{a^2} + \frac{24 a' w_0'}{a^3} - \frac{8 w_2}{a^2}  + \frac{w_0''''}{2} - \frac{6 a'^2 w_0''}{a^2}   \nonumber & \\
&  - \frac{12 a'^3 w_0'}{a^3} + 3 w_2'' + \frac{14 a' w_2'}{a} + \frac{12 a'^2 w_2}{a^2} + \frac{8}{a^3} \frac{\partial}{\partial z} \left[ \sigma  + \frac{(\Bou_{\kappa} \kappa_s - \Bou_{\mu} \mu_s)}{a} \biggl((a w_0)' \right.  \nonumber \\
& \left.    + \frac{(a^3 w_2)' - a'^2 (a w_0)' - (a a' (a w_0)')' }{2} - \frac{\mathcal{C}}{2} \left( (a^2 w_0)' + \frac{(a^4 w_2)'}{4} - \frac{a'^2 (a^2 w_0)'}{2} \right) \biggr) \right]   \nonumber & \\
& + \frac{8 \Bou_{\mu} \mu_s a'}{a^3} \left( \frac{a w_2'}{4} + 2 a'' w_0' \right) + \frac{8 \Bou_{\mu}}{a^4} \frac{\partial}{\partial z} \biggl[ \mu_s a ( 2 w_0' + (a^2 w_2)' - 4  a'^2 w_0' - a a' w_0'') \biggr]  \nonumber & \\
& + \frac{8 \Bou_{\mu} \mu_s a' w_0'}{a^4}  + \frac{8 \Bou_{\mu} a'^2 (\mu_s a w_0')'}{a^4},
\end{align}
\begin{align}\label{eq:secondorder_surf}
& \frac{\partial \Gamma}{\partial t} + \frac{(a w_0 \Gamma)'}{a} + \frac{(a^3 w_2 \Gamma)' - a'^2 (a w_0)' \Gamma - (a a' (a w_0)')' \Gamma}{2 a}  \nonumber & \\
& - \frac{\mathcal{C} \, \Gamma}{2 a}  \left( (a^2 w_0)' + \frac{(a^4 w_2)'}{4} - \frac{a'^2 (a^2 w_0)'}{2} \right) = 0.
\end{align}
Note that equations~\eqref{eq:secondorder_continuity}--\eqref{eq:secondorder_surf}, together with the three equations of state $\sigma(\Gamma)$, $\mu_s(\Gamma)$ and $\kappa_s(\Gamma)$, form a closed system which determines the temporal evolution of $a(z,t)$, $w_0(z,t)$, $w_2(z,t)$ and $\Gamma(z,t)$.

\subsection{Temporal stability analysis of the 1D models}\label{subsec:1DLSA}

Performing numerical simulations of the temporal evolution of the liquid thread using the 1D models derived above is beyond the scope of the present work. Hence, as a first step, it is interesting to check the validity of both 1D approximations in the linear regime by comparing their temporal stability properties with those provided by the full axisymmetric dispersion relation derived in~\S\ref{sec:results}. Linearising equations~\eqref{eq:1Dcontinuity}--\eqref{eq:1Dsurfactant2} and~\eqref{eq:secondorder_continuity}--\eqref{eq:secondorder_surf}, and expanding in normal modes, the dispersion relations associated with the leading-order model and the parabolic model are straightforwardly obtained, as follows.
\begin{enumerate}
\item Leading-order model:
\begin{equation}\label{eq:1D_DR}
\frac{2 \omega^2 \Reycap}{k^2} - 1 + k^2 + \beta + \omega( 6  + 9 \Bou_{\mu} + \Bou_{\kappa}) = 0.
\end{equation}
\item Parabolic model:
\begin{align}\label{eq:1D_Parabolic_DR}
& \Reycap \omega^3(16 + 4 k^2) + \omega^2[128 + k^2(96 + 40 \Bou_{\kappa} + 104 \Bou_{\mu}) + k^4(18 + 9 \Bou_{\kappa} + 25 \Bou_{\mu})] \nonumber & \\
& + \omega k^2 [-8 + 40 \beta + 384 \, \Reycap^{-1} \, (1 + 3\Bou_{\mu}/2 + \Bou_{\kappa}/6) + k^2 (7 + 9 \beta + \Reycap^{-1} \, (96 + 80 \Bou_{\kappa} \nonumber & \\
& + 144 \Bou_{\mu} + 64 \Bou_{\kappa}\Bou_{\mu})) + k^4 (1+\Reycap^{-1} \, (14 + 9 \Bou_{\kappa} + 25 \Bou_{\mu}))] + \Reycap^{-1} \, k^2 [64(\beta-1) \nonumber & \\
& + k^2(48 + 80 \beta - 16 \Bou_{\kappa} - 16 \Bou_{\mu} + 64 \beta \Bou_{\mu}) + k^4(15 + 9 \beta + 16 \Bou_{\kappa} + 16 \Bou_{\mu}) + k^6] \nonumber & \\
& - \frac{16 \beta}{\Reycap \, \omega}k^4 (1- k^2) = 0.
\end{align}
\end{enumerate}

\begin{figure}
\includegraphics[width=1\textwidth]{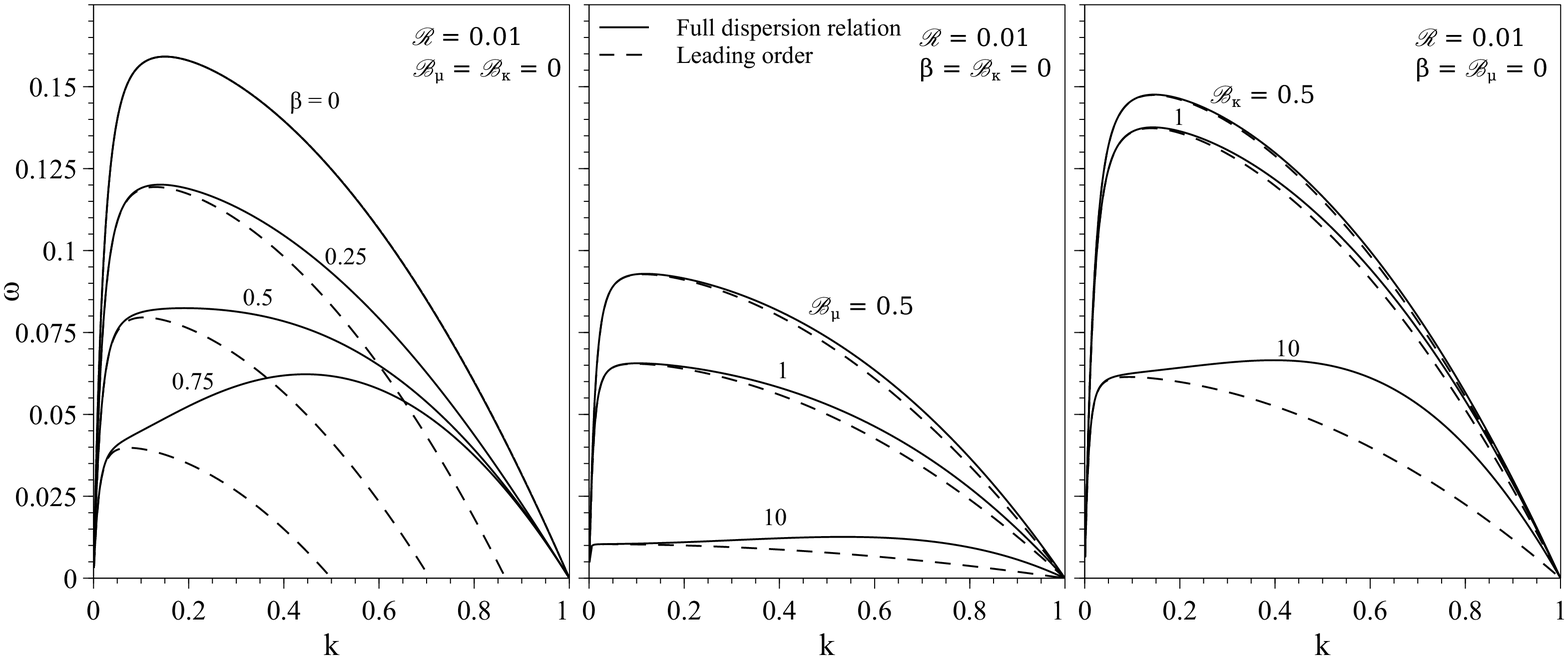}
\caption{\label{fig:figure7} Comparison between the full axisymmetric dispersion relation~\eqref{eq:dimless_DR} (solid lines) and the dispersion relation associated with the leading-order 1D model~\eqref{eq:1D_DR} (dashed lines) for $\Reycap = 0.01$ and varying the value of $\beta$ ($a$), $\Bou_{\mu}$ ($b$) and $\Bou_{\kappa}$ ($c$). The values of the parameters are indicated near each curve.}
\end{figure}

\begin{figure}
\includegraphics[width=1\textwidth]{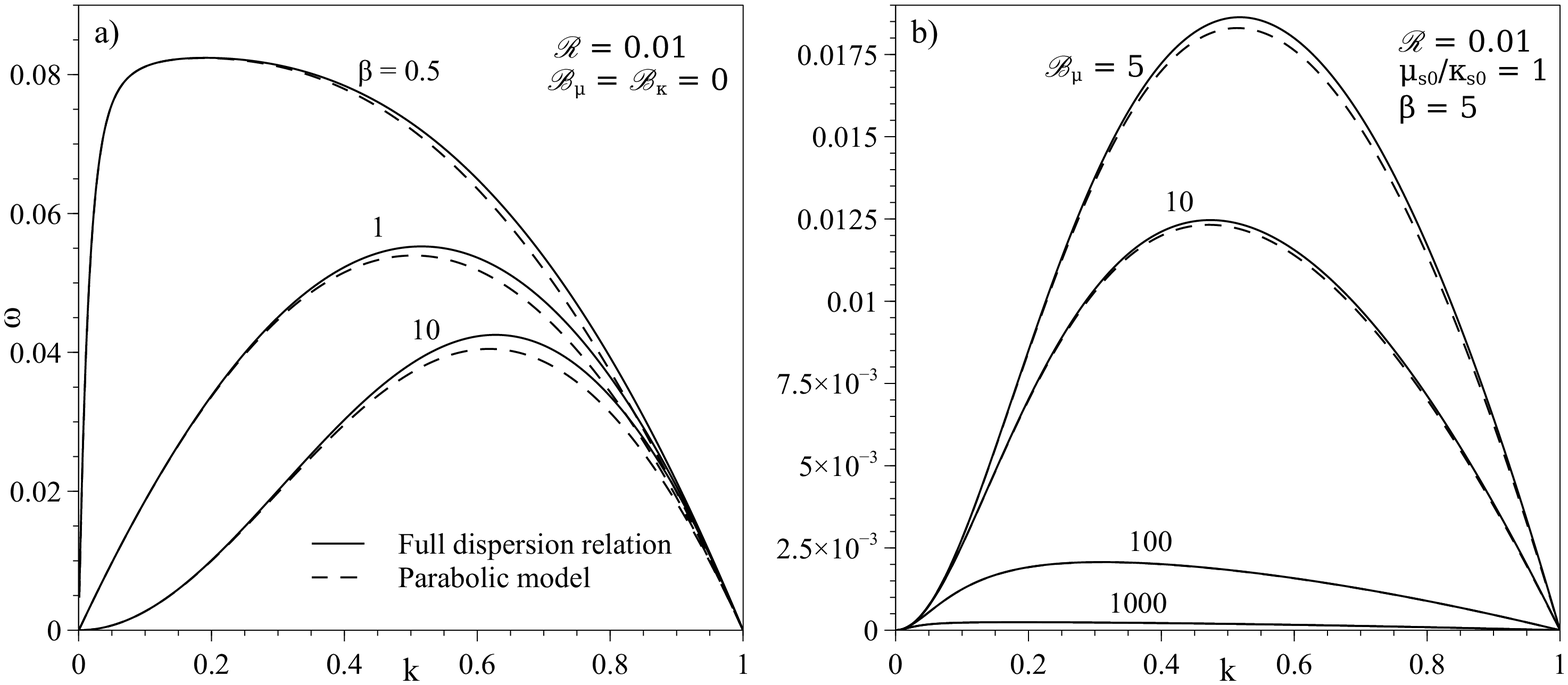}
\includegraphics[width=1\textwidth]{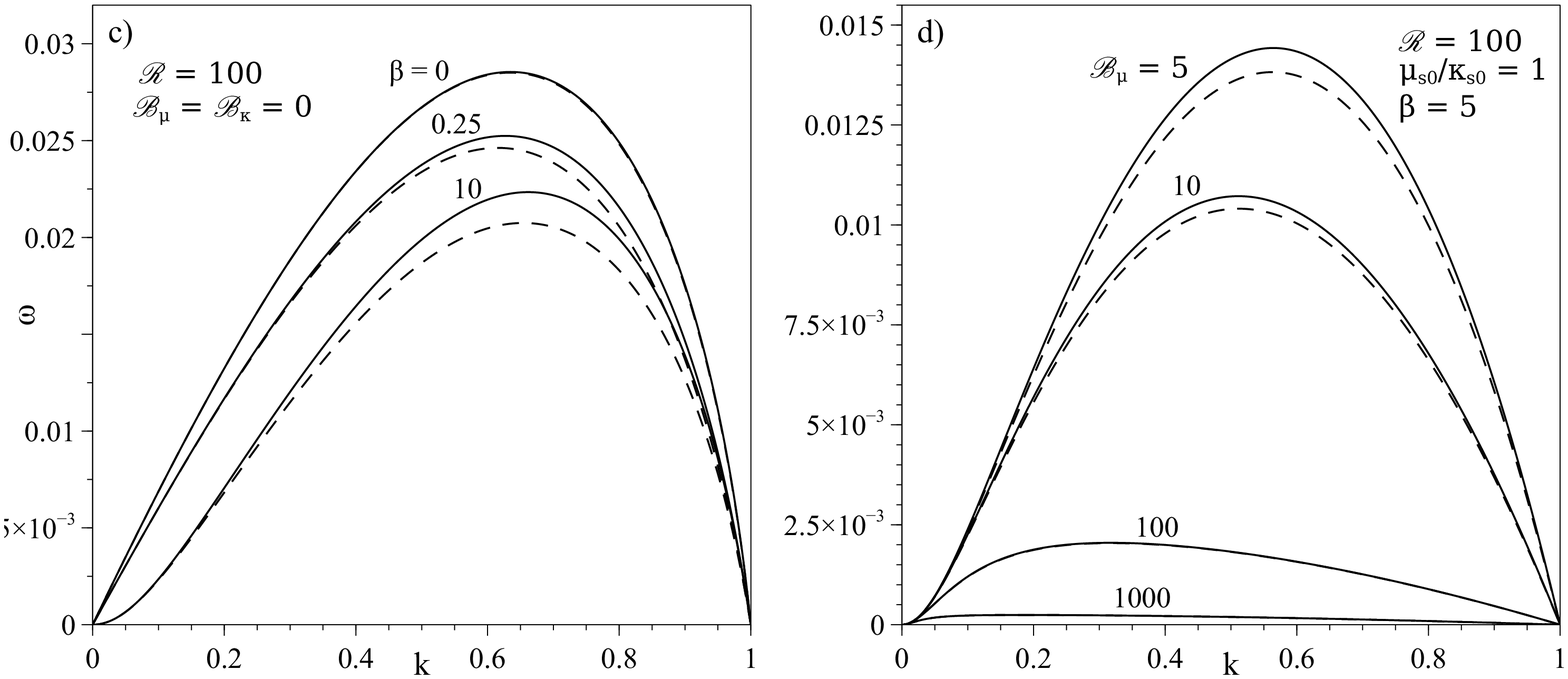}
\caption{\label{fig:figure8} Comparison between the full axisymmetric dispersion relation~\eqref{eq:dimless_DR} (solid lines) and the dispersion relation associated to the parabolic 1D model~\eqref{eq:1D_Parabolic_DR} (dashed lines) for $\Reycap = 0.01$ ($a$,$b$) and $\Reycap = 100$ ($c$,$d$). Panels ($a$,$c$) show the amplification curves for $\Bou_{\mu} = \Bou_{\kappa} = 0$ and several values of $\beta$, and ($b$,$d$) those for $\beta = 5$, $\mu_{s0}/\kappa_{s0} = 1$ and several values of $\Bou_{\mu}$ indicated near each curve.}
\end{figure}

To illustrate the performance of the two different 1D models deduced in the present work, figures~\ref{fig:figure7} and~\ref{fig:figure8} compare, respectively, the amplification curves obtained with the leading-order model~\eqref{eq:1D_DR} and with the parabolic model~\eqref{eq:1D_Parabolic_DR}, with those given by the full axisymmetric dispersion relation~\eqref{eq:dimless_DR}, for several values of $\beta$, $\Bou_{\mu}$ and $\Bou_{\kappa}$. In the case of the elasticity parameter, $\beta$, figure~\ref{fig:figure7}(a) shows that the leading-order 1D model fails at predicting the linear behaviour of the liquid column as $\beta$ increases, as was already pointed out by~\citet{Timmermans02}. In fact, according to this 1D approximation, the liquid column is stable for every value of $k$ if $\beta \geq 1$. With regard to $\Bou_{\mu}$ and $\Bou_{\kappa}$, figure~\ref{fig:figure7}($b$,$c$) shows that the leading-order 1D approximation also fails at predicting the growth rate $\omega(k)$, especially when $\Bou_{\mu} \gg 1$ and $\Bou_{\kappa} \gg 1$.

In contrast, the parabolic model captures the linear regime with high accuracy for every value of $\beta$, $\Bou_{\mu}$ and $\Bou_{\kappa}$, as evidenced by the results of figure~\ref{fig:figure8}. The agreement between both $\omega(k)$ curves is slightly worse when $\beta \to \infty$, whereas it improves when $(\Bou_{\kappa}, \Bou_{\mu}) \gg 1$ independently of the value of $\beta$. In particular, we have checked that in the limit $\Bou_{\mu}\to\infty$, the dispersion relation~\eqref{eq:1D_Parabolic_DR} is identical, at leading order, to equation~\eqref{eq:Bmugg} deduced from the exact dispersion relation~\eqref{eq:dimless_DR}.

It is worth mentioning that~\citet{Timmermans02} also derived a higher-order approximation usually known as the averaged-parabolic model, but only accounting for gradients of $\sigma$ and not surface viscosities. These authors demonstrated that, for a very viscous thread, i.e. $\Reycap \to 0$, the averaged-parabolic model shows a very good agreement with the full expression of the dispersion relation~\eqref{eq:dimless_DR} for $\Bou_{\mu} = \Bou_{\kappa} = 0$ and different values of $\beta$. Indeed, the accuracy of their averaged-parabolic model is slightly better than that of the parabolic model developed herein in the limit $(\Bou_{\mu},\,\Bou_{\kappa}) \to 0$, when only surface elasticity is considered. Note that, in the case of a clean interface, a similar conclusion was obtained by~\citet{GyC}, who found that the 1D averaged-parabolic model performs slightly better than the 1D parabolic model in predicting the linear amplification curve.

In summary, the leading-order 1D equations are suitable for small values of $\beta$, $\Bou_{\mu}$ and $\Bou_{\kappa}$, including the limit of a clean interface~\citep{EggersDupont,GyC}. However, when $(\beta, \Bou_{\mu}, \Bou_{\kappa}) \gtrsim 1$, a higher-order approximation like the parabolic model derived herein, or the averaged-parabolic model due to~\citet{Timmermans02}, is needed to obtain good quantitative results in the linear regime.

\section{Conclusions \label{sec:conclusions}}

We have provided the full expressions for the normal and tangential stress boundary conditions at a general axisymmetric interface coated with insoluble surfactant, with surface viscosity effects correctly taken into account in the Boussinesq--Scriven approximation~\citep{Boussinesq1913,Scriven60}. These boundary conditions have been applied to obtain the dispersion relation between growth rate and wavenumber for the canonical case of the capillary instability of a free uniform thread of Newtonian liquid coated with an insoluble monolayer. Through a modal temporal analysis we have shown that the maximum growth rate decreases monotonically with the surface shear and dilatational viscosities, while the corresponding wavenumber displays a more complex non-monotonic dependence. 

In addition, we have presented two different 1D approximations that account for the effect of surface viscoelasticity due to the presence of insoluble surfactants, namely a leading-order model and a second-order parabolic one. By comparing the dispersion relations associated with these 1D models with the exact one, it is deduced that the validity of the leading-order model is limited to small enough values of the elasticity parameter and of the shear and dilatational Boussinesq numbers. In contrast, the parabolic model provides a dispersion relation in close agreement with the temporal amplification curves of the exact dispersion relation for the whole range of dimensionless parameters. 

Although a nonlinear analysis of surface viscous effects, including their influence on the breakup of the thread and on the formation of satellite drops, is beyond the scope of the present study, it clearly deserves further investigation. To that end, the parabolic 1D model developed herein could well be a very useful tool to obtain the nonlinear time evolution of the liquid thread at a reduced computational cost compared with numerical simulations of the full conservation equations.

\begin{acknowledgments}

The authors thank the Spanish MINECO, Subdirecci\'on General de Gesti\'on de Ayudas a la Investigaci\'on, for its support through projects DPI2014-59292-C3-1-P, DPI2015-71901-REDT and DPI2017-88201-C3-3-R. These research projects have been partly financed through European funds. A.M.-C. also acknowledges support from the Spanish MECD through the grant FPU16$/$02562. The authors are grateful to Prof. Benoit Scheid for suggesting the use of the lubrication approximation, and to an anonymous referee for making important contributions that improved the present work.

\end{acknowledgments}



\begin{appendix}
\section{Details on the derivation of the interfacial stresses} \label{app:diffgeom}
For completeness, several terms of equation~\eqref{eq:eqsurface} of the main text are deduced in detail:
\begin{align}
& \bnabla_s \bcdot \boldm{u}_s = \frac{1}{\sqrt{1+a'^2}} \frac{\partial \boldm{u}_s}{\partial z} \bcdot \boldm{t} + \frac{1}{a} \frac{\partial \boldm{u}_s}{\partial \theta} \bcdot \boldm{e}_{\theta} = \frac{(a u_t)'}{a \sqrt{1+a'^2}} + \frac{u_n}{a \sqrt{1+a'^2}} - \frac{a'' u_n}{(1+a'^2)^{3/2}} \nonumber & \\
& = \frac{1}{\sqrt{g}} \frac{\partial (\sqrt{\mathsfi{g}_{22}} u_t ) }{\partial z} + \mathcal{C} u_n = \bnabla_s \cdot \boldm{u}_t + \mathcal{C} u_n,
\end{align} 
\begin{align}
(\bnabla_s \boldm{u}_s)_{11} = \frac{1}{\sqrt{1 + a'^2}}\frac{\partial \boldm{u}_s}{\partial z} \bcdot \boldm{t} = \frac{u_t'}{\sqrt{1+a'^2}} - \frac{a'' u_n}{(1+a'^2)^{3/2}},
\end{align}
\begin{align}
(\bnabla_s \boldm{u}_s)_{13} = \frac{1}{\sqrt{1 + a'^2}}\frac{\partial \boldm{u}_s}{\partial z} \bcdot \boldm{n} = \frac{u_n'}{\sqrt{1+a'^2}} + \frac{a''  u_t}{(1+a'^2)^{3/2}},
\end{align}
\begin{align}
(\bnabla_s \boldm{u}_s)_{22} = \frac{1}{a}\frac{\partial \boldm{u}_s}{\partial \theta} \bcdot \boldm{e}_{\theta} = \frac{a' u_t}{a \sqrt{1+a'^2}} + \frac{u_n }{a\sqrt{1+a'^2}},
\end{align}
\begin{align}
(\bnabla_s \boldm{u}_s)_{12} = (\bnabla_s \boldm{u}_s)_{21} = (\bnabla_s \boldm{u}_s)_{23} = (\bnabla_s \boldm{u}_s)_{31} = (\bnabla_s \boldm{u}_s)_{32} = (\bnabla_s \boldm{u}_s)_{33} = 0,
\end{align}
\begin{align}
& \bnabla_s \bcdot \left\{ \mu_s \left[(\bnabla_s \boldm{u}_s) \bcdot \mathsfbi{I}_s +\mathsfbi{I}_s \bcdot (\bnabla_s \boldm{u}_s)^{\text{T}} \right]\right\}  = \left\{ \frac{1}{a} \frac{\partial}{\partial{z}} \left[\frac{2 \mu_s a}{1+a'^2} \left( u_t' - \frac{a'' u_n}{1+a'^2} \right) \right] \nonumber \right. & \\
& \left. - \frac{2 \mu_s a'}{1+a'^2} \left[ \frac{a' u_t + u_n}{a^2} - \frac{a''}{1+a'^2} \left(u_t' - \frac{a'' u_n}{1+a'^2} \right)  \right] \right\} \boldm{t} \nonumber & \\
& -\frac{2 \mu_s}{1+a'^2} \left[\frac{a' u_t + u_n }{a^2} - \frac{a''}{1+a'^2} \left(u_t' -\frac{a'' u_n}{1+a'^2} \right)  \right] \boldm{n}
\end{align}


\section{Matrix of the homogeneous linear system} \label{app:matrix_entries}

The substitution of normal-mode variables evaluated at $r = R$ in~\eqref{eq:bc_modes_normal} and~\eqref{eq:bc_modes_tangential} provides a homogeneous linear system for the constants of integration $A$ and $B$ of the form $\mathsfbi{M} \bcdot \boldsymbol{\phi} = 0$, where $\boldsymbol{\phi} = (A,B)^T$ and the following four entries of $\mathsfbi{M}$:

\begin{align}
& \mathsfi{M}_{11}= I_0(kR)\left[1 + \frac{k^2}{\rho \omega} \left( 2 \mu - \frac{E}{\omega R}-\frac{k_{s0}-\mu_{s0}}{R} \right) \right]  \nonumber & \\ & + \frac{k I_1 (kR)}{\rho \omega R} \left[\frac{E - \sigma_0(1-k^2 R^2)}{\omega R} - 2 \mu + \frac{\kappa_{s0} + \mu_{s0}}{R} \right],
\end{align}
\begin{align}
& \mathsfi{M}_{12}= \tilde{k} I_0(\tilde{k} R)\left[\frac{E}{\omega R} - 2\mu +\frac{k_{s0}-\mu_{s0}}{R}\right]  \nonumber & \\
& + \frac{I_1 (\tilde{k}R)}{R} \left[2 \mu +\frac{\sigma_0(1-k^2 R^2) - E}{\omega R}   -  \frac{\kappa_{s0} + \mu_{s0}}{R}\right],
\end{align}
\begin{equation}
\mathsfi{M}_{21}= \frac{i k^2}{\rho \omega} \left[-k I_0(k R)\left(\frac{E}{\omega} + k_{s0}+\mu_{s0} \right) +  I_1 (k R) \left(\frac{E}{\omega R} - 2 \mu  + \frac{\kappa_{s0} - \mu_{s0}}{R} \right)\right],
\end{equation}
\begin{align}
\mathsfi{M}_{22}= i \left[ k \tilde{k} I_0(\tilde{k} R)\left(\frac{E}{\omega} + k_{s0}+\mu_{s0}\right)   + I_1 (\tilde{k}R) \left(\frac{\mu \tilde{k}^2}{k} + \mu k - \frac{k E}{\omega R} - \frac{k(\kappa_{s0} - \mu_{s0})}{R} \right) \right].
\end{align}

\end{appendix}

\bibliographystyle{jfm}

\end{document}